\documentclass[10pt]{article}
\usepackage{amsmath}
\usepackage{amssymb}
\usepackage{graphicx}
\usepackage{color} 
\usepackage{cite}
\usepackage{color}
\usepackage{indentfirst} 
\usepackage{url} 
\usepackage{booktabs} 
\usepackage{subfigure} 
\usepackage{xr} 
\usepackage[labelfont=bf,labelsep=period,justification=raggedright]{caption}

\makeatletter
\renewcommand{\@biblabel}[1]{\quad#1.}
\makeatother

\date{}

\begin{document}

\begin{flushleft}
{\Large
\textbf{Spatio-temporal Dynamics of Foot-and-Mouth Disease Virus in South America}
}
\\
Luiz Max Carvalho$^{1,2\ast}$,
Nuno Rodrigues Faria$^{3,4}$,
Andres M.~Perez$^{5}$,
Marc A.~Suchard$^{6,7}$
Philippe Lemey$^{3}$,
Waldemir de Castro Silveira$^{8}$,
Andrew Rambaut$^{1,9,10}$,
Guy Baele$^{3}$
\\
\bf{1} Institute of Evolutionary Biology, University of Edinburgh, Edinburgh, United Kingdom.\\
\bf{2} Scientific Computing Program, Oswaldo Cruz Foundation, Rio de Janeiro, Brazil.\\
\bf{3} Department of Microbiology and Immunology, Rega Institute -- KU Leuven, Leuven, Belgium.\\
\bf{4} Department of Zoology, University of Oxford, Oxford, United Kingdom.\\
\bf{5} Department of Veterinary Population Medicine, University of Minnesota, St. Paul, United States of America.\\
\bf{6} Departments of Biomathematics and Human Genetics, David Geffen School of Medicine at UCLA, University of California, Los Angeles,  United States of America.\\
\bf{7} Department of Biostatistics, UCLA Fielding School of Public Health, University of California, Los Angeles,  United States of America.\\
\bf{8} Research and Development Division, Trimatrix Applied Biotechnology Ltd, Rio de Janerio, Brazil.\\
\bf{9}  Fogarty International Center, National Institutes of Health, Bethesda, MD,  United States of America.\\
\bf{10} Centre for Immunology, Infection and Evolution at the University of Edinburgh, Edinburgh, United Kingdom.
$\ast$ E-mail: lm.carvalho@ed.ac.uk
\end{flushleft}
\section*{Abstract}
Although foot-and-mouth disease virus (FMDV) incidence has decreased in South America over the last years, the pathogen still circulates in the region and the risk of re-emergence in previously FMDV-free areas is a veterinary public health concern.
In this paper we merge environmental, epidemiological and genetic data to reconstruct spatiotemporal patterns and determinants of FMDV serotypes A and O dispersal in South America.
Our dating analysis suggests that serotype A emerged in South America around 1930, while serotype O emerged around 1990.
The rate of evolution for serotype O  was significantly higher compared to serotype A.
Phylogeographic inference identified two well-connected sub networks of viral flow, one including Venezuela, Colombia and Ecuador; another including Brazil, Uruguay and Argentina.
The spread of serotype A was best described by geographic distances, while trade of live cattle was the predictor that best explained serotype O spread.
Our findings show that the two serotypes have different underlying evolutionary and spatial dynamics and may pose different threats to control programmes.

Key-words: Phylogeography, foot-and-mouth disease virus, South America, animal trade.


\section*{Introduction}

Foot-and-mouth disease virus (FMDV) is a rapidly evolving picornavirus and the causative agent of foot-and-mouth disease (FMD), the most important disease of domestic and wild cloven-hoofed animals~\cite{review}.
The virus can be classified in seven serotypes, three of which (A, O, and C) have circulated in South America.
Serotype A caused large epidemics throughout the Southern cone in recent years~\cite{Perez2001, Malirat2012}, while endemic circulation has been mostly limited to Venezuela~\cite{Malirat2012}.
Historically, serotype O has been the most prevalent serotype on the continent, but is now limited to areas in the Andean region, and in particular to Ecuador~\cite{andean}.
Serotype C on the other hand was last encountered in the continent in $1995$ in Brazil~\cite{review_eradication}.
Historical reports suggest that FMDV arrived in South America in the late years of the 19th century with European colonization~\cite{Naranjo2013, tully}. 
By the 1970s, FMD was widespread in the region, with several large-scale epidemics being caused by multiple subtypes~\cite{Saraiva2003}.
In South America, FMD control and eradication has traditionally been pursued using a combination of mass vaccination programs~\cite{vaccinationSA} and control of animal movements from areas in which FMDV infection was suspected.
Over time, passive and active surveillance programs have, with different degrees of success, managed the early detection of FMDV.
In order to achieve complete eradication however, the strains involved in epidemics - especially those in previously FMDV-free areas - need to be accurately characterised.

Phylogenetic analyses have proven useful in recovering the transmission pathways from genetic data~\cite{cottam2007, cottam2008} and providing insight into the processes that drive re-emergence~\cite{combining}.
More recently, molecular epidemiology tools have been used to infer the origin and evolutionary history of emerging strains in South America~\cite{Perez2001, Malirat2007, andean, Malirat2011, Maradei2013}.
However, as pointed out by Di Nardo, Knowles \& Paton~\cite{combining}, a common feature of FMDV molecular epidemiology studies is that  joint evaluation of epidemiological, environmental and genetic data has usually been performed outside of an unified quantitative framework.
In the face of many sources of information, ranging from genetic data to environmental data on host distribution and outbreak counts, it's desirable to have a framework capable of integrating these sources of information coherently.
Phylodynamics combines population genetics and epidemiology to explicitly  model the interaction between ecological processes such as migration and selection and the shape of the phylogenies~\cite{grenfell, vphylodynamics}.
Bayesian phylodynamics offers an attractive statistical framework to combine multiple sources of information while marginalizing over the topology space, thus accommodating phylogenetic uncertainty.
In particular, phylogeographic methods can be employed to understand viral spatial dynamics under explicit spatial diffusion models~\cite{roots}.
Further, an important research goal is to gain insight into the major determinants of FMDV spread in the continent.
Since animal movements constitute a major threat to eradication programs~\cite{movements}, using animal trade data as predictors can be a valuable tool to understand the role of livestock commerce in the spread of FMDV.
For example, Nelson et al. ~\cite{Nelson2011} coupled swine trade data and genetic data to show that swine movements in the United States drove the spread of a novel influenza virus of the H1 subtype.

Here, we investigate the phylodynamic patterns of serotypes A and O in South America using all publicly available VP1 (1D) sequences for those serotypes in South America, sampled over a long time-period (1955-2010 for serotype A and 1994-2010 for serotype O) in nearly all south American countries affected by FMD.
We apply Bayesian phylogeographic methods to investigate the evolutionary dynamics of serotypes A and O in South America incorporating  genetic, spatial and epidemiological data such as livestock trade, geographic distances and vaccination coverage.
This flexible Bayesian phylogeographic framework allows for the testing of hypotheses concerning viral dispersal, while naturally accommodating phylogenetic uncertainty~\cite{roots, towards}.
We use BEAST~\cite{beast2012} to infer time-structured phylogenies and reconstruct past population dynamics, to which we overlay vaccination and serotype-specific notification data.
To study the factors driving re-emergence, we use data on livestock trade and geographical distances as predictors for viral spatial diffusion and compare competing spatial dynamics models involving each predictor using recently developed methods. 

\section*{Results}

\subsection*{Evolutionary rates and times of origin of FMDV serotypes A and O circulating strains}

First, we perform model selection using path-sampling (PS) and stepping-stone (SS) to estimate marginal likelihoods and calculate (log) Bayes factors (BF) in order to select the best fitting demographic (tree) model and molecular clock model.
Our results favor a non-parametric skyride model, that allows fluctuations in demographic growth through time, over the constant population size assumption (serotype A: log BF = $11$; serotype O: log BF = $21$).
Using a similar approach, we  find decisive support for the relaxed clock over the strict molecular clock model.
Particularly, the exponential relaxed molecular clock provides a better fit for serotype O data (log BF = $56$) while the log-normal relaxed molecular clock model provides a better fit for serotype A (log BF = $34$) (see Table~\ref{stab:treeclockselection} for details). 
The coefficient of variation for the log-normal relaxed molecular clock model for serotype A has a posterior mean of $0.32$ (95\% highest posterior density [HPD] interval: $0.21$--$0.43$) indicating substantial rate heterogeneity among lineages in the phylogeny.
Root-to-tip plots against sampling times from maximum likelihood phylogenies constructed with PhyML~\cite{phyml} suggest a linear trend (Figure~\ref{sfig:root-to-tip}), which encourages the pursuit of more detailed estimates.
In addition, we compare a model with and without dated-tips (the `temporal signal' test~\cite{Faria2012, Baele2012}, see Methods) and find significant temporal structure for both serotypes (A: log BF = $310$; O: log BF = $348$), showing that sufficient temporal information is embedded in the sequence data of both serotypes under investigation, which is essential for the estimation of divergence times and the reconstruction of the population dynamics in natural time units~\cite{MEP}. 

\begin{figure}[!ht]
\begin{center}
\subfigure[A]{\includegraphics[scale=.45]{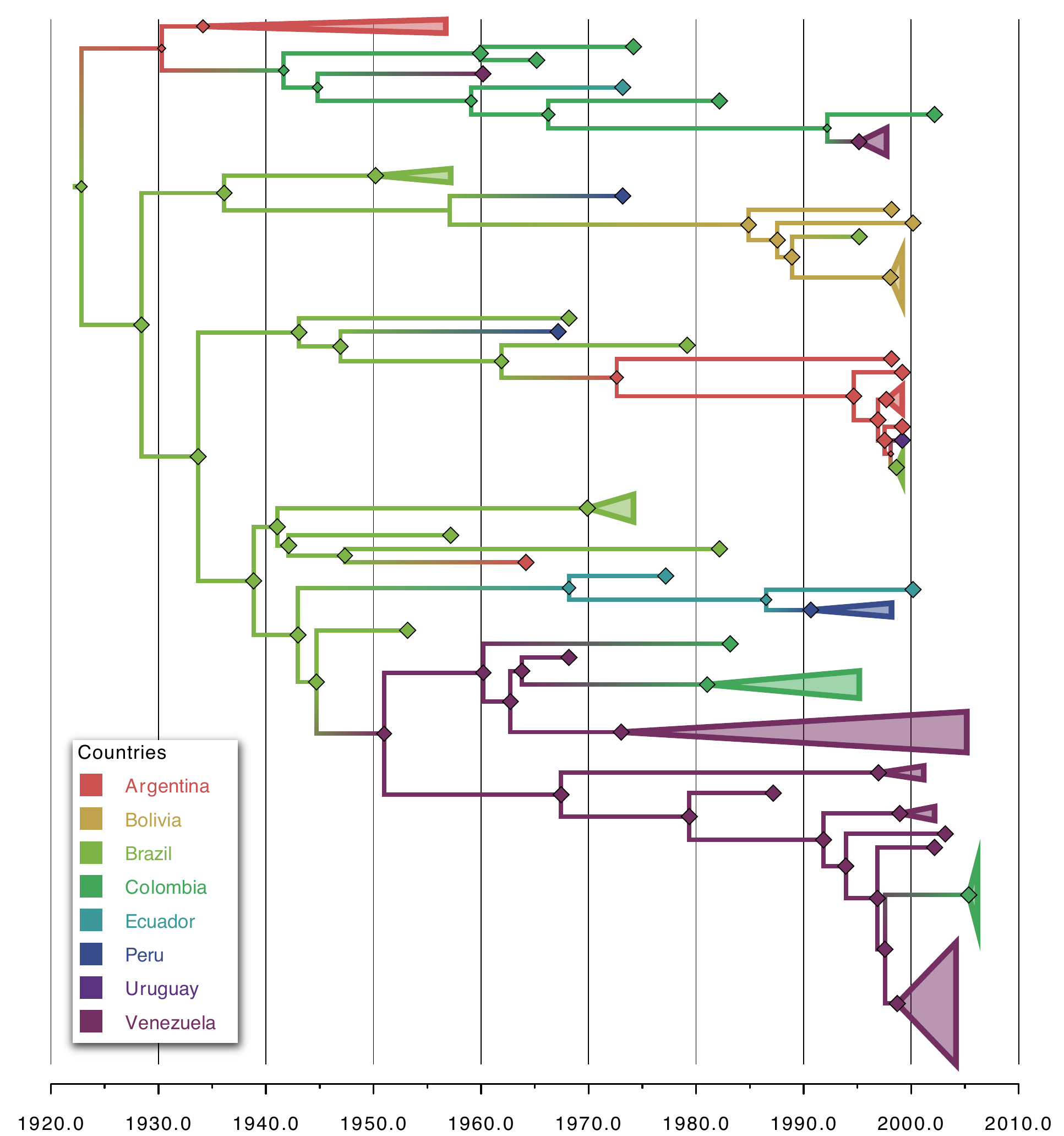}}\\
\subfigure[O]{\includegraphics[scale=.30]{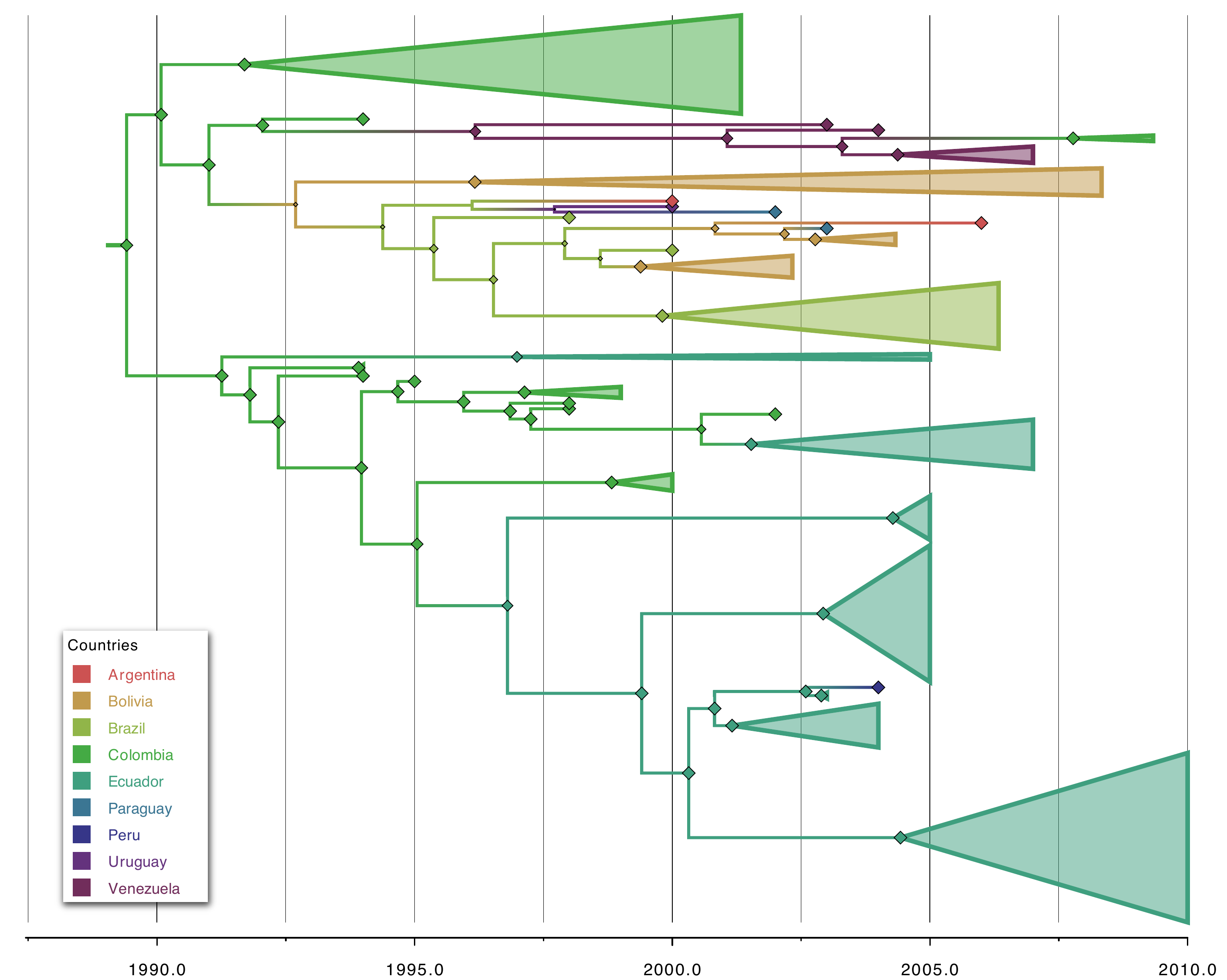}}
\end{center}
\caption{}
\label{fig:trees}
\end{figure}

Figure~\ref{fig:trees} shows the estimated maximum clade credibility (MCC) trees for serotype A and serotype O.
The time to most recent common ancestor (tMRCA) of the circulating serotype A strains is estimated at $1932$ ($95\%$ HPD: $1925$--$1939$) and  $1989$ ($95\%$ HPD: $ 1986$--$1991$) for serotype O, indicating a more recent origin of the latter. 
Using the combination of best fitting demographic and molecular clock models for each serotype, the evolutionary rate for serotype A is estimated at $\approx 4 \times 10^{-3}$ substitutions/site/year.
The estimated evolutionary rate for serotype O is approximately 2.5 times faster than serotype A, i.e.~at $\approx 1 \times 10^{-2}$, consistent with previous studies~\cite{tully, Carvalho2013, Muellner2011}.
We refer to Supplementary Text S2, Tables~\ref{stab:SB_A} and~\ref{stab:SB_O} for more detailed estimates.

Figure~\ref{fig:trees}A shows that sequences from the same country tend to cluster in small clades, although the inferred phylogenies for both serotypes also show considerable interspersing of lineages, indicating trans-border FMDV spread.
For example, the Argentinian serotype A sequences are grouped in two clades, that either comprise only Argentinian isolates or include sequences from Brazil and Uruguay (Figure~\ref{fig:trees}A).
Interestingly, the majority of the isolates from Venezuela and Colombia fall together within two distinct clades.
For serotype O there is also interspersing of clades from different locations, with one large clade featuring interleaved Ecuadorian and Colombian clades (Figure~\ref{fig:trees}B).
A smaller clade of Colombian isolates are found interspersed within Venezuelan isolates.
In addition, isolates from Ecuador are grouped with isolates from Colombia, suggesting that the intra-country dynamics of FMDV between these two countries is intrinsically linked.
These observations indicate that there is spatial structure in the data, i.e. that there is association between phylogeny and geography.
We further explore this using Bayesian Tip-Significance testing (BaTS)~\cite{bats} to calculate phylogeny-trait association indices such as the parsimony score (PS), association index (AI) and maximum clade size (MC) (see Text S2 for details).
Both serotypes show significant (randomised p-value $< 0.05$) location-tip association for the three indices (Table~\ref{stab:BaTS}).
As an overall summary, our data sets present a high degree of spatial signal, justifying richer phylogeographic analyses to study the transmission network of FMDV on the continent.

\begin{center}
 [Figure~\ref{fig:trees} about here]
\end{center}

\subsection*{Spatial Dynamics of FMDV in South America}

\begin{figure}[H]
\begin{center}
\subfigure[A -- $1945$ ]{\includegraphics[scale=.20]{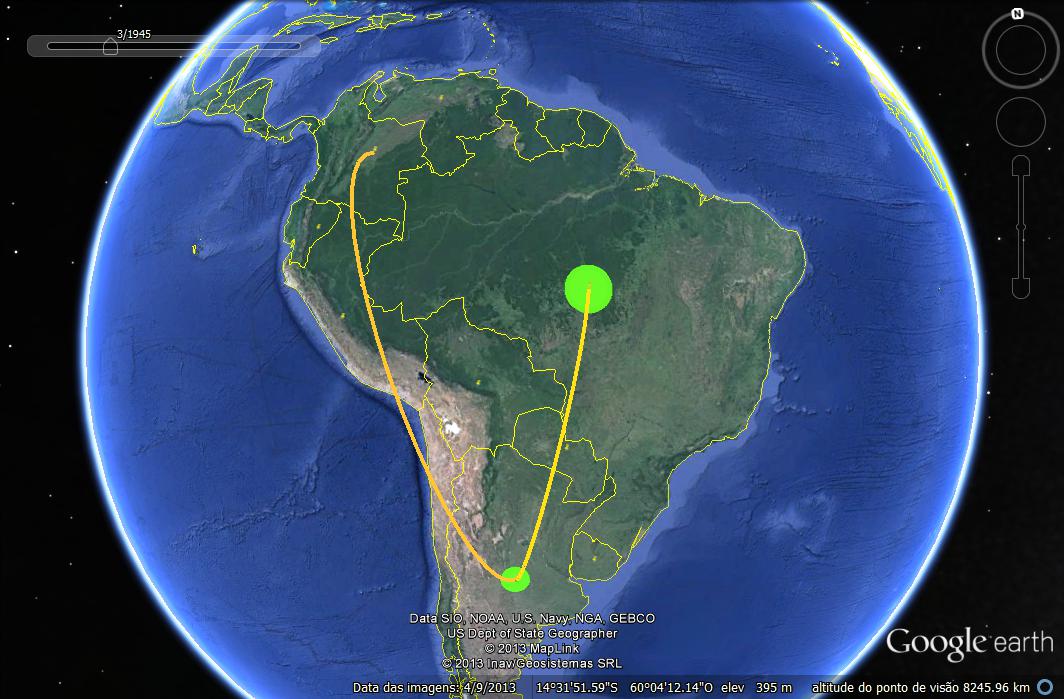}}
\subfigure[O -- $1995$ ]{\includegraphics[scale=.20]{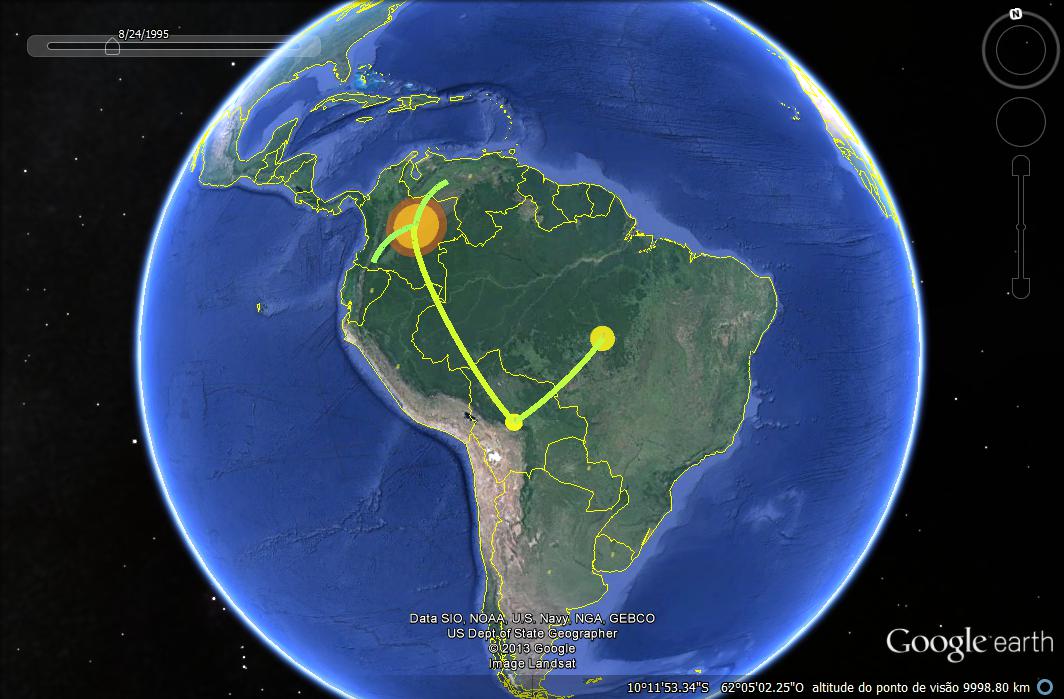}}\\
\subfigure[A -- $1965$ ]{\includegraphics[scale=.20]{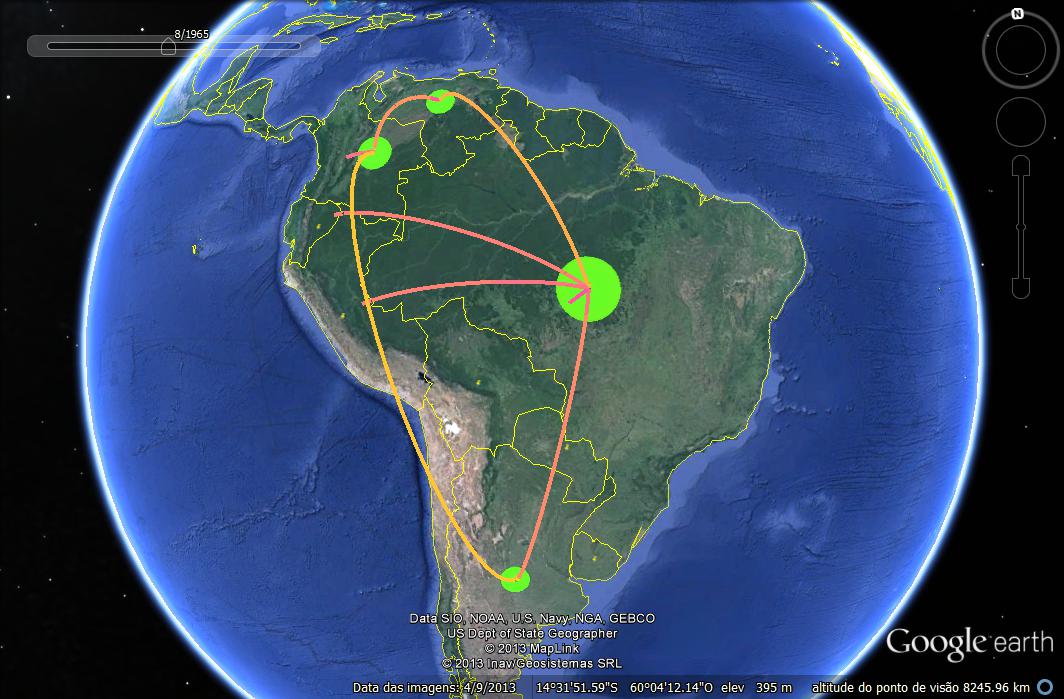}}
\subfigure[O -- $2000$ ]{\includegraphics[scale=.20]{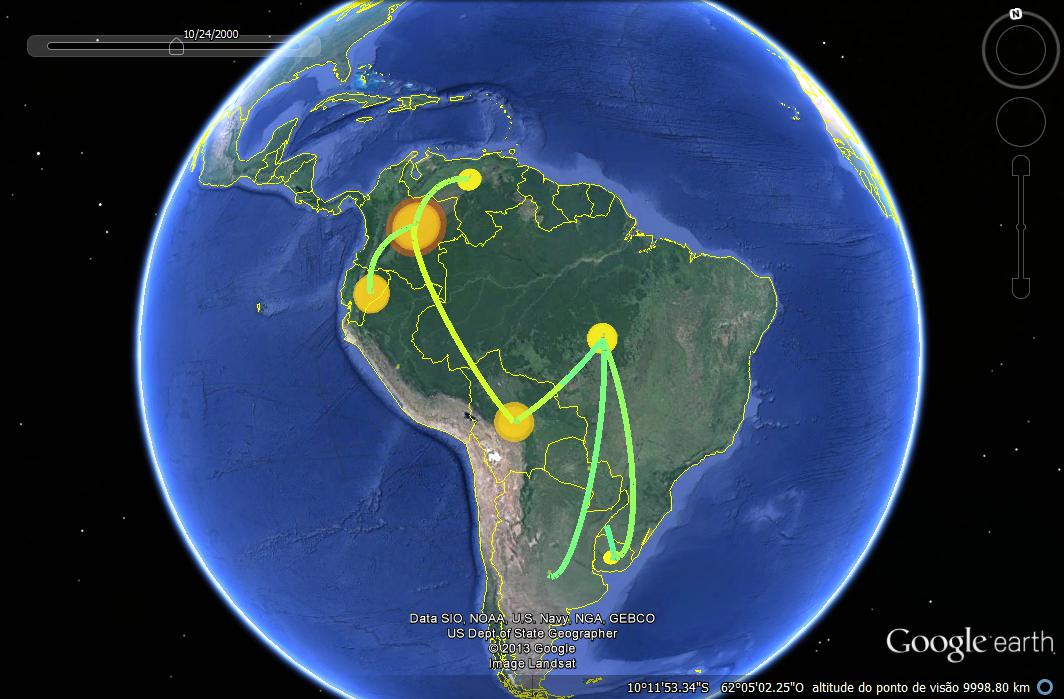}}\\
\subfigure[A -- $1980$ ]{\includegraphics[scale=.20]{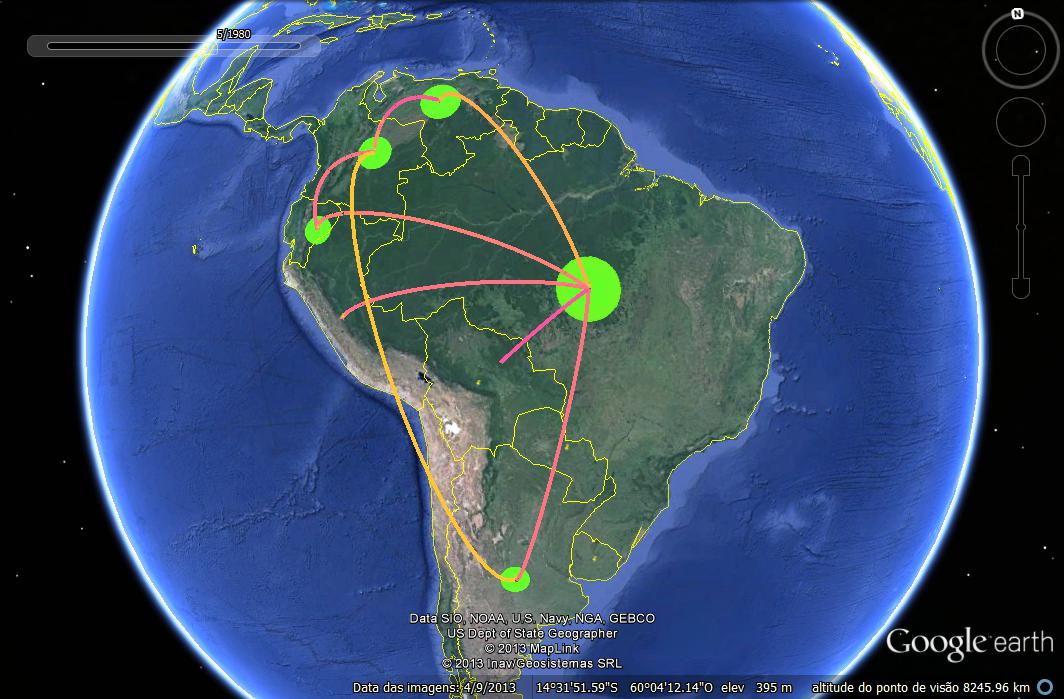}}
\subfigure[O -- $2005$ ]{\includegraphics[scale=.20]{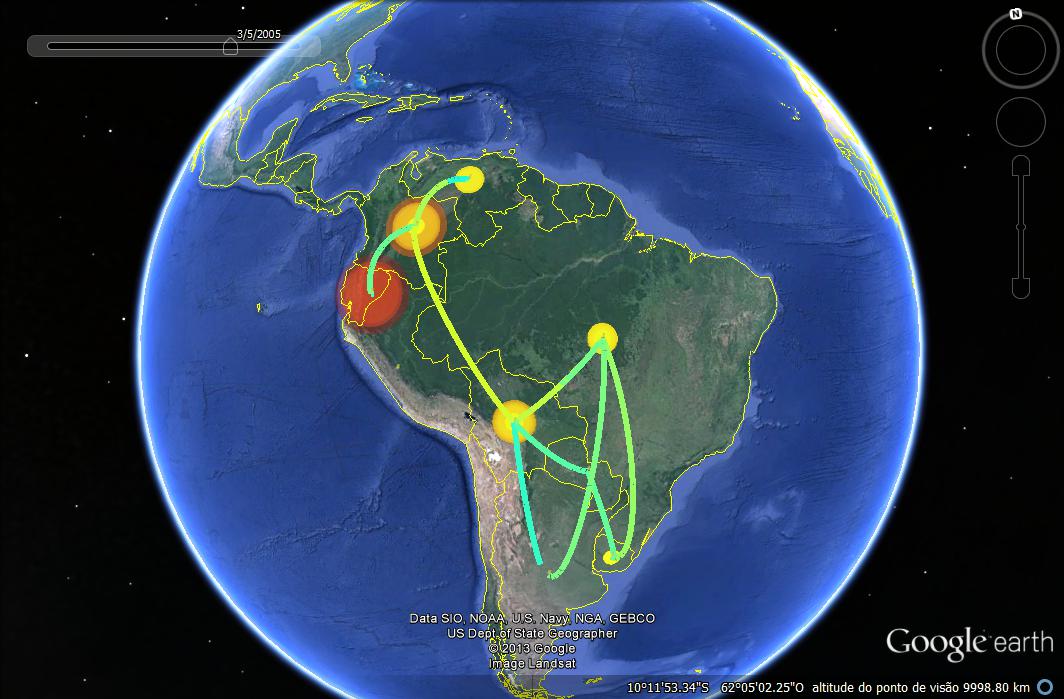}}\\
\subfigure[A -- $2008$ ]{\includegraphics[scale=.20]{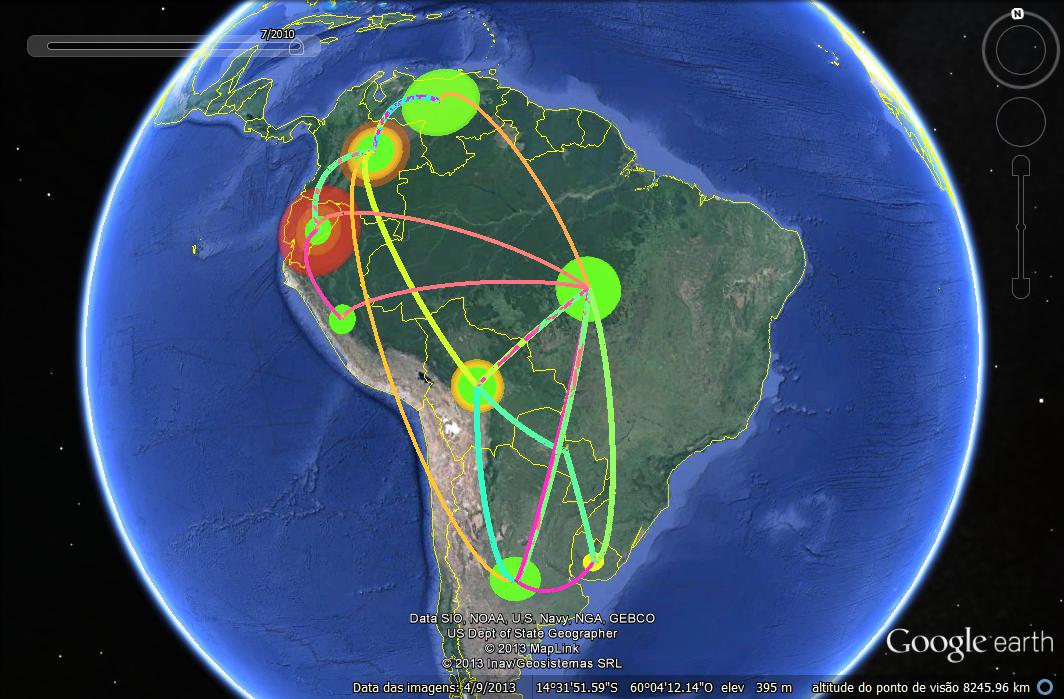}}
\subfigure[O -- $2010$ ]{\includegraphics[scale=.20]{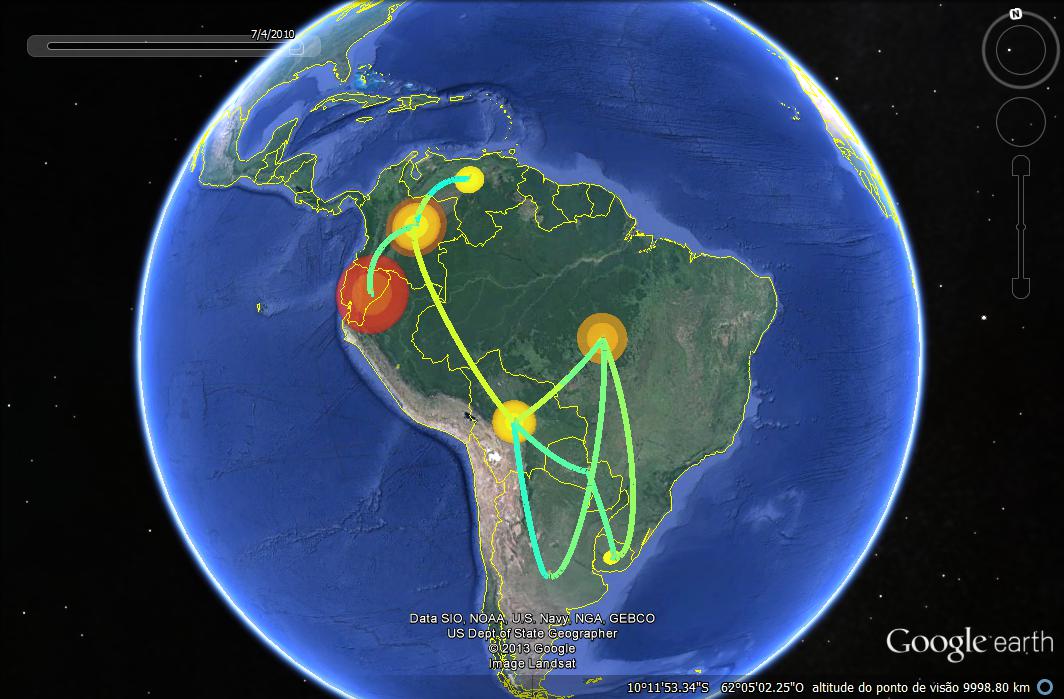}}
\end{center}
\caption{}
\label{fig:migration}
\end{figure}

To gain insight into the spatio-temporal process of FMDV spread, we employ an asymmetric continuous-time Markov chain (CTMC) phylogeographic model~\cite{roots} implemented in BEAST~\cite{beast2012}, coupled with model averaging using a Bayesian stochastic search variable selection (BSSVS) procedure.
Spatial projections of the maximum clade credibility trees (Figure~\ref{fig:migration}) show that Brazil and Colombia are the most strongly connected regions (hubs) both for serotypes A and O, respectively. 
In the serotype A phylogeography, Brazil is connected to Argentina, Uruguay, Venezuela and Colombia, while for serotype O, Colombia is connected to Venezuela, Ecuador and Bolivia.
These results suggest different spatial patterns for the two serotypes.

We observe that for serotype A mainly long-range migration routes are inferred for the period before $1945$.
This may however result from the limited amount of data available before $1970$, which limits accurate inference migration events in the past (see \textbf{Discussion}).
Further, there is a major expansion in spatial spread from $1945$ to $1965$, characterized by Brazil as a source of virus for the rest of the continent.
We observe a slower spread, mainly through short range routes, in the period $1965-1980$.
The $1980-2008$ time window is characterised by FMDV serotype A flow into Peru and Paraguay and the increase of intra-country diversity (depicted by the radius of the displayed circles in Figure~\ref{fig:migration}).   

On the other hand, the serotype O expansion seems to have occurred in the mid $1990$'s.
Up to 1995, our reconstruction suggests that Colombia and Brazil appear to act as primary and secondary viral sources, respectively.
From 1995 to $2000$, whilst Colombia acted as the main source for the northern/Andean regions of the continent, the spread to the Southern Cone seems to stem from Brazil.
The period $2000-2010$ is characterized by a decrease in viral dispersal movements (almost no new edges added to the network) concomitant with an increase in viral diversity within countries, specially Ecuador, Colombia and Brazil.

We subsequently turn our attention to the statistical support for epidemiological linkage between pairs of locations using Bayes factors~\cite{roots}.
For serotype A we find high statistical support for viral migration between Uruguay and Argentina (BF = $985$), Venezuela and Colombia (BF = $812$), Bolivia and Brazil (BF = $25$) and Brazil and Argentina (BF = $6$).
On the other hand, for serotype O the most supported links are found between Colombia and Ecuador (BF = $557$), Ecuador and Peru (BF = $223$), Venezuela and Colombia (BF = $52$), Paraguay and Argentina (BF = $6$). 

\begin{center}
 [Figure~\ref{fig:migration} about here]
\end{center}

\begin{figure}[H]
\begin{center}
\includegraphics[scale=.90]{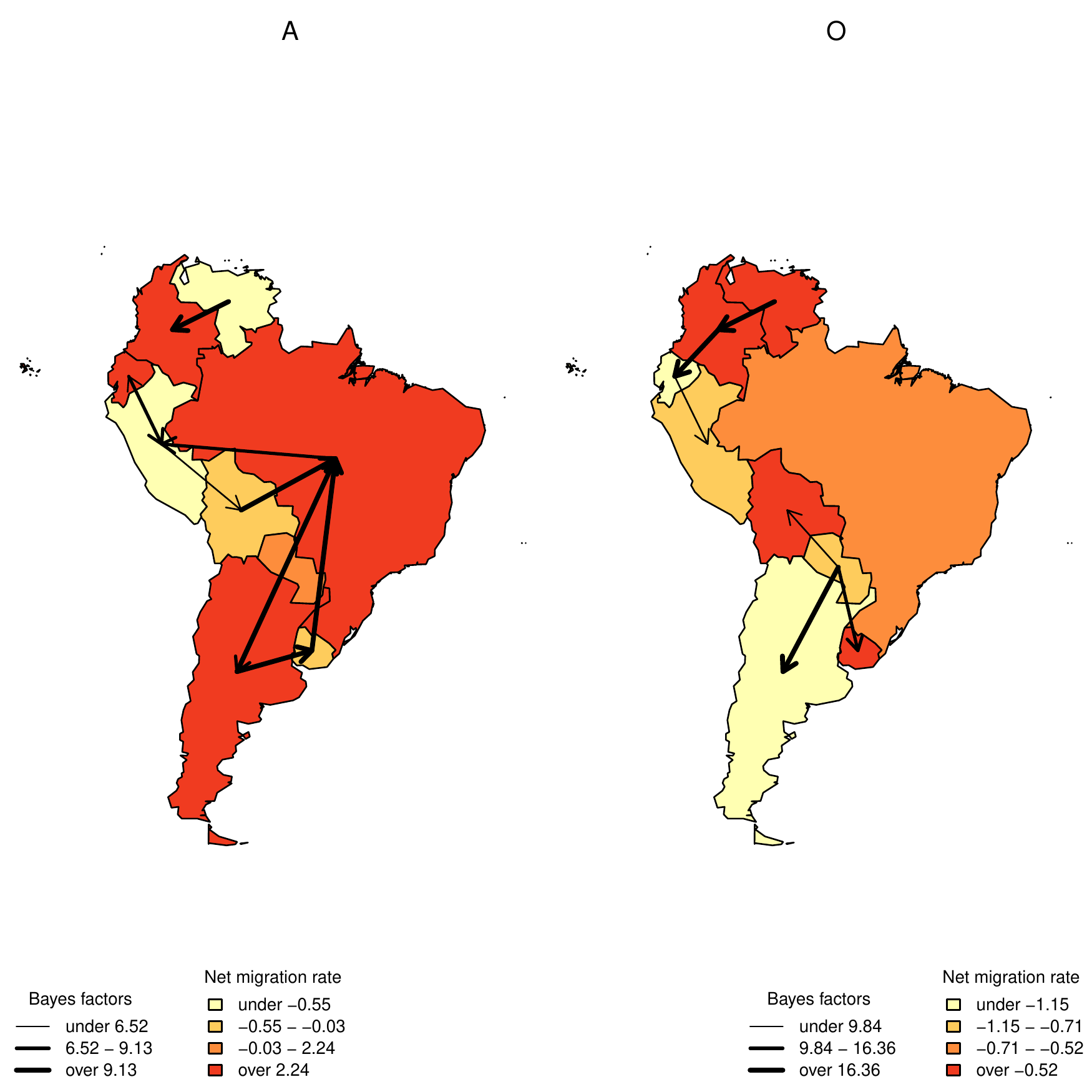}
\end{center}
\caption{}
\label{fig:mj&BFs}
\end{figure}

To infer which countries have most significantly contributed to the dispersal of FMDV serotypes, we estimate net migration flow (efflux - influx) using a robust counting approach~\cite{Minin2008}.
Figure~\ref{fig:mj&BFs} shows that the supported migration routes differ for each serotype, with some overlapping routes connecting Venezuela and Colombia. 
We also observe the pattern of emitters/receivers to be different for each serotype.
While for serotype A Brazil and Argentina act mainly as viral exporters, Venezuela and Bolivia present the highest net rates for serotype O.
Overall, net migration rates are substantially higher for serotype A in comparison to serotype O (Figure~\ref{fig:mj&BFs}).
The presence of Bolivia as a viral hub for serotype O dispersal and its negligible role for serotype A  dispersal is an interesting difference between the two serotypes. 
Likewise, Brazil is highly connected in the serotype A network but not on the serotype O network.

\subsubsection*{Preferential Spread of FMDV across neighbouring countries}

We compare the transition counts between countries that share borders and those that do not and find that for serotype A the posterior median transition count between neighboring countries amounts to $22$ (95 \% HPD: $17$--$27$) while ``non-border'' transitions have a posterior median of $4$ (95~\% HPD: $0$--$7$). 
For serotype O, we observe similar results, with $16$ ($13$--$20$) and $1$ ($0$--$3$) location-transitions being observed for border (B) and non-bordering (NB) countries, respectively.
Considering each South American country as a possible discrete state, there are $72$ possible transitions ($34$ B and $38$ NB transitions).
For serotype A, these numbers have to be adjusted because we have samples from $8$ countries, resulting in $\text{B} = 26$ and $\text{NB} = 30$.
Remarkably, however, the importance of long-range migration routes seems to differ for both serotypes, since the posterior median proportion of ``non-border'' transitions is considerably different (A: $0.14$, $0.00$--$0.28$  and O: $0.05$, $0.00$--$0.20$).

\begin{center}
 [Figure~\ref{fig:mj&BFs} about here]
\end{center}

\subsubsection*{Spatial origins for main South American FMDV outbreaks}

\begin{figure}[H]
\begin{center}
\subfigure[][]{\includegraphics[scale=.40]{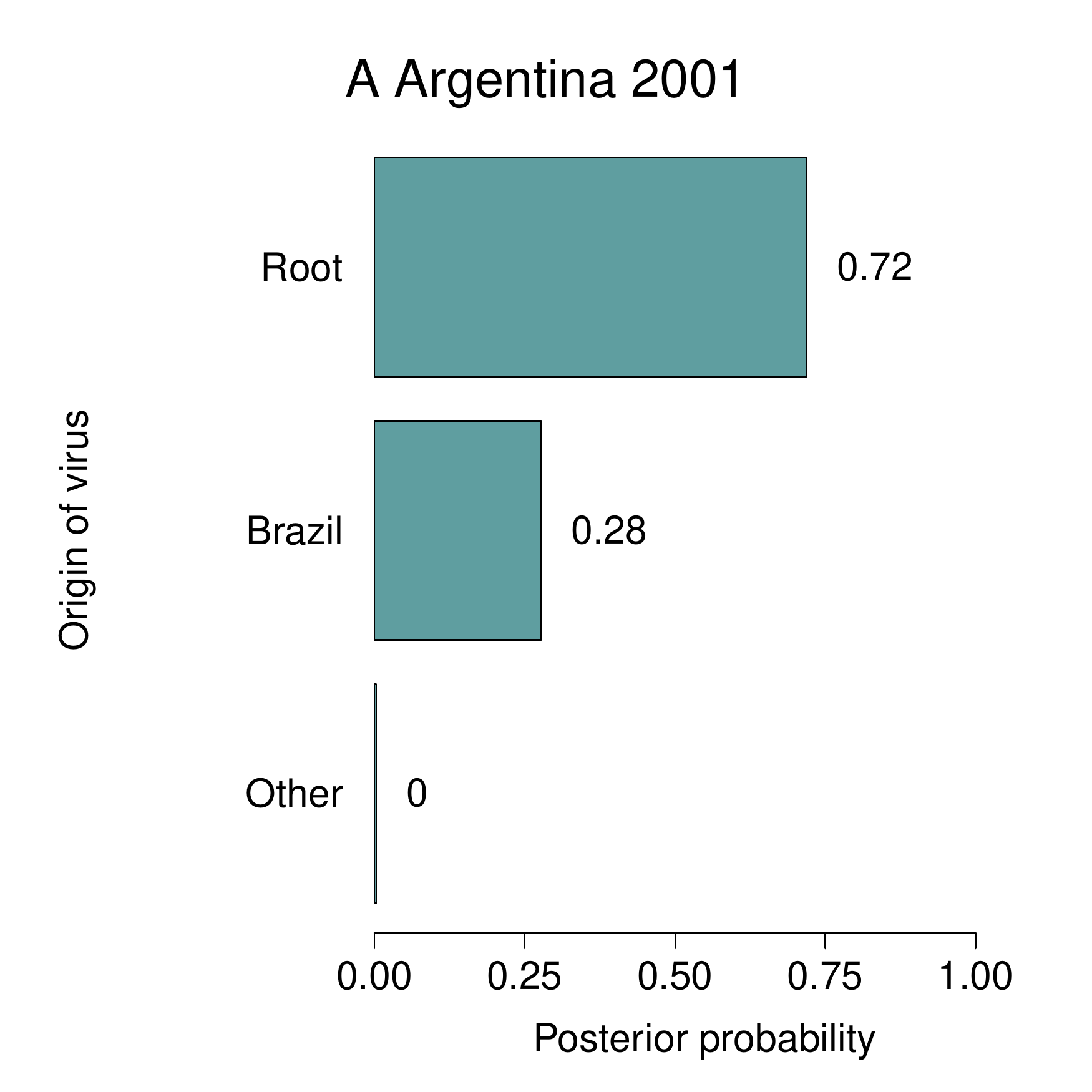}}
\subfigure[][]{\includegraphics[scale=.40]{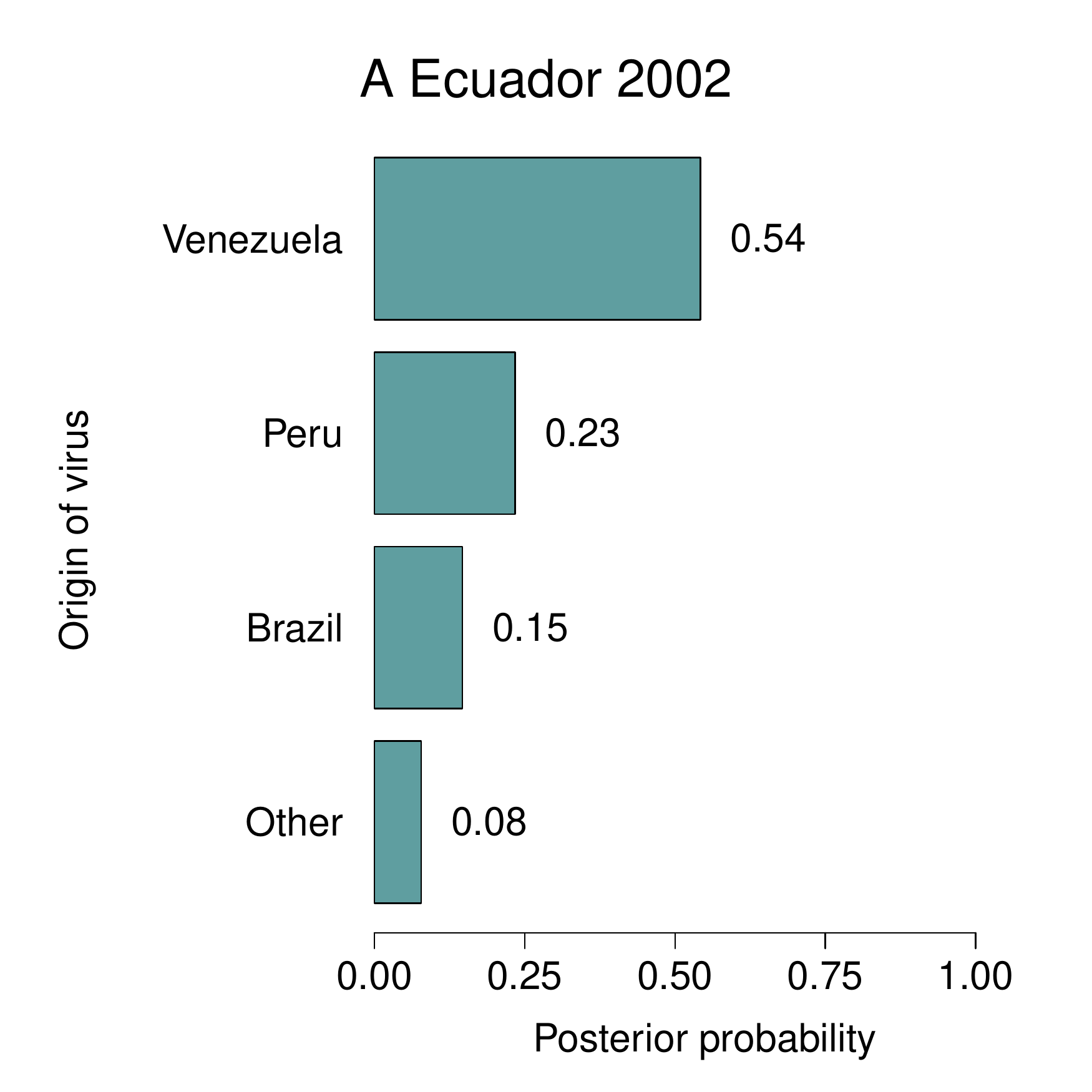}}\\
\subfigure[][]{\includegraphics[scale=.40]{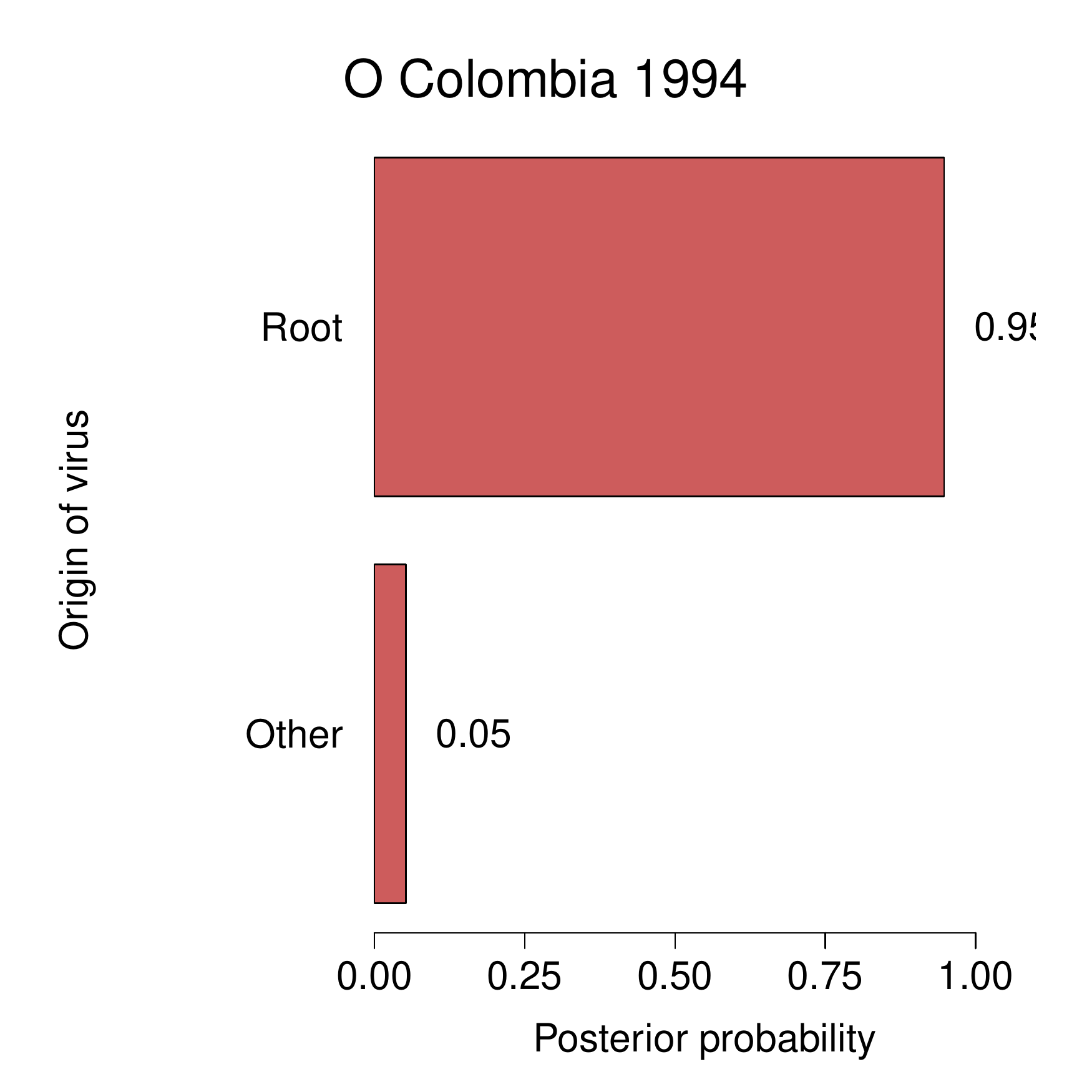}}
\subfigure[][]{\includegraphics[scale=.40]{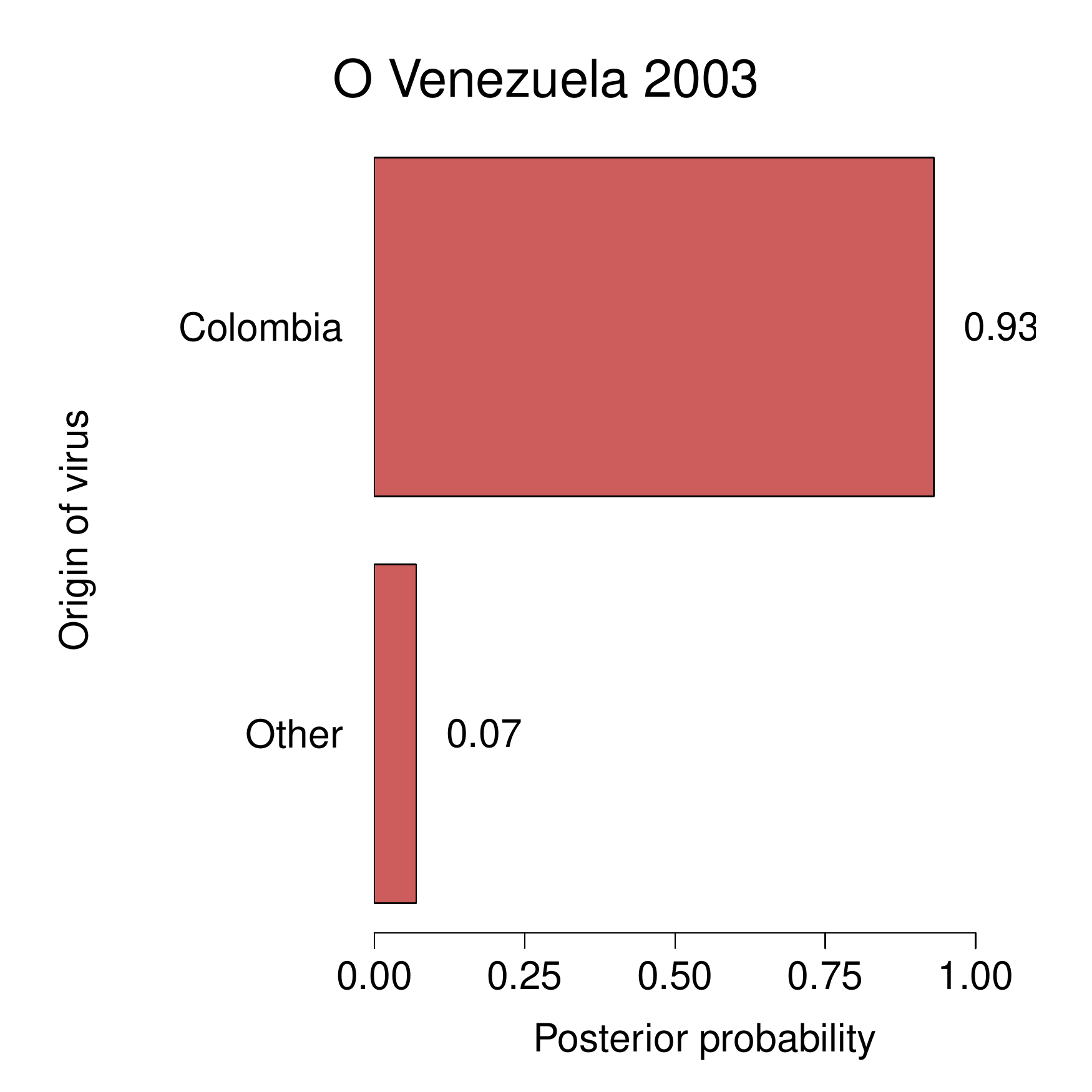}}
\end{center}
\caption{}
\label{fig:epidemictracing}
\end{figure}

Figure~\ref{fig:epidemictracing} shows the posterior distribution of the location of origin for some epidemics of interest.
The most probable origin of the strains isolated in Argentina $2001$ (Figure~\ref{fig:epidemictracing}A) was Argentina (posterior probability = $0.72$) while Brazil received a posterior probability of $0.28$.
Results for the Brazilian strains of the same year point to Argentina as the source of the epidemic with high probability, confirming the strong link between the two countries (Figure~\ref{sfig:epitrac}B).
However, these results should be interpreted with caution due to the low number of sequences from Uruguay.
Contrasting to the connectivity with Argentina, Bolivia $2001$ seems to have had an independent introduction from Peru, as shown in Figure~\ref{sfig:epitrac}C
We provide evidence of Venezuela as a major viral source for serotype A in its region in Figure~\ref{fig:epidemictracing}B, where we show the origins of the Ecuadorian strain isolated in $2002$.
This notion of Venezuela as a seeder is further strengthened by the fact that the Colombian $2008$ strain was most likely imported from Venezuela (posterior probability $\approx 1$, shown in Figure~\ref{sfig:epitrac}D).

Similar to what was found for Venezuela and serotype A, Colombia seems to be the source of most of the strains circulating in the northern part of South America.
As with the introduction of serotype A in Colombia in $2008$, the strains from Venezuela $2003$ have a high probability of being imported from Colombia (Figure~\ref{fig:epidemictracing}D).
Consistent with these findings, we show in Figures~\ref{sfig:epitrac}B and~\ref{sfig:epitrac}C that Colombia was the most probable origin of the strains in Ecuador ($2002$).
The link between Venezuela and Colombia thus seems to be supported from data for both serotypes, although the main viral seeder varies for each serotype (see \textbf{Discussion}).

\begin{center}
 [Figure~\ref{fig:epidemictracing} about here]
\end{center}

To address the relevance of epidemiological predictors, such as geographic distance and livestock trade on viral diffusion, we collected data on the trade of live cattle, pigs and sheep  as well as geographic distances, and estimate marginal likelihoods for each of these predictors.
The sub-network connecting Venezuela, Colombia and Ecuador is present in all three networks, as are the connections between the Southern Cone countries.
Additionally, for the sheep and cattle networks, we identify some long-range trade routes, for instance the trade of sheep between Argentina and Colombia.
We use geographic distances as well as the data on the trade of livestock to specify CMTC rate matrix priors (potential predictors), and we employ PS/SS to calculate (log) marginal likelihoods for each predictor to determine the importance of each variable for viral spread~\cite{Carvalho2013, Nelson2011}.
For serotype A, we identify geographic distance as the best predictor for viral diffusion between countries (log BF$\approx 4$, compared to the equal-rates gamma prior), whereas the exchange of cattle is considered to be the best predictor for serotype O spread (log BF$\approx 13$). 
Further, while we find moderate statistical support for geographic distances as predictors of viral spread for serotype A, this predictor has higher statistical support for serotype O (log BF$\approx 9$).

\begin{center}
 [Table~\ref{tab:preds} about here]
\end{center}

For each predictor, we also assess the location state distribution at the root. 
This vector contains the posterior probability of each country being the origin of the circulating strains.
In Table~\ref{tab:roots} we show that for serotype A there is discordance between predictors about which country was the most probable source of FMDV on the continent.
Peru is considered the most probable location of origin according to the `cattle' predictor, but notably not according to the best fitting predictor (geographic distance), for which Argentina is estimated as the origin with moderate probability ($0.75$).
Argentina is also found to be the most probable origin according to the equal-rates gamma prior model ($Pr(\text{root})=0.84$).
For the swine trade predictors, Colombia is found to be the spatial origin.
For serotype O, the predictors show much more concordance, and Colombia is ascertained to be the spatial viral origin for all predictors with high probability (Table~\ref{tab:roots}).
Regarding spatial signal extraction, as measured by the Kullback-Liebler (KL) divergence (relative  to a uniform prior) at the root~\cite{roots} (see also Text S2) for each predictor, we find KL divergences ranging from $3.86$ to $5.91$, showing highly concentrated distributions at the root. 
The predictors with highest KL divergences are pigs and sheep for serotypes A and O respectively.
The results for serotype A are therefore inconclusive since the most efficient predictor (pigs) point to a different origin than the best fitting predictor (distance).
Further studies are necessary to address the spatial origins of FMDV serotype A in South America.

\begin{center}
 [Table~\ref{tab:roots} about here]
\end{center}
\subsubsection*{Sensitivity analysis of spatial sampling heterogeneity}

Since our data sets contain unbalanced geographic samples (Table~\ref{stab:reps}), we conduct a detailed sensitivity analysis to assess the robustness of our results to over-representation of some locations~\cite{Faria2012, fluPNAS, Bedford2010, polar}.
To this end, we randomly draw five sequence sub-samples for each serotype in which over-represented geographic regions are down-sampled and run the asymmetric CTMC (with BSSVS) analysis for each sub-sample.
We calculate log Bayes factors and observe good agreement between sub-samples in terms of supported routes for both serotypes (see Figures~\ref{sfig:bssvsA} and~\ref{sfig:bssvsO}). 

To further investigate the agreement between sub-samples, i.e. if parameter estimates are consistent across subsamples, we also compute the $L_1$ matrix distance norm across the estimated posterior mean rate matrices for each sub-sample, from which we detect no aberrant samples (see Text S2 for detailed results).
Another aspect of interest is the amount of information extracted from each sample, which we measure by calculating Kullback-Leibler~\cite{KL} divergence between prior and posterior distributions of spatial location at root~\cite{roots}.
Tables~\ref{stab:ED_A} and~\ref{stab:ED_O} show that for all sub-samples in both serotypes, there is lower information extraction when compared to the full analysis, with moderately concentrated posterior distributions at root.
This result is expected however, because each subsample has less data than the full data set.
For serotype O, inference about location of origin is consistent across samples, with Venezuela being the most probable country of origin.
In the case of serotype A however, there is some disagreement concerning the most probable root state, with Argentina being pointed as root by one of the sub-samples.
Further, we observe very similar support for Venezuela as the root for serotype O when compared to Brazil as the root for serotype A.
While serotype O presents an averge probability of $0.47$ that Venezuela was the root, the average probability that Brazil was the root for serotype A is $0.50$.
We take the lack of agreement between the distribution at root for the sub-samples and the results from the full data to be the result of the lack of information in the down-sampled analyses, as evidenced by the much lower KL divergences (Tables~\ref{stab:ED_A} and~\ref{stab:ED_O}). 

\subsection*{Demographic reconstruction of FMDV}

\begin{figure}[H]
\begin{center}
\includegraphics[scale=1.0]{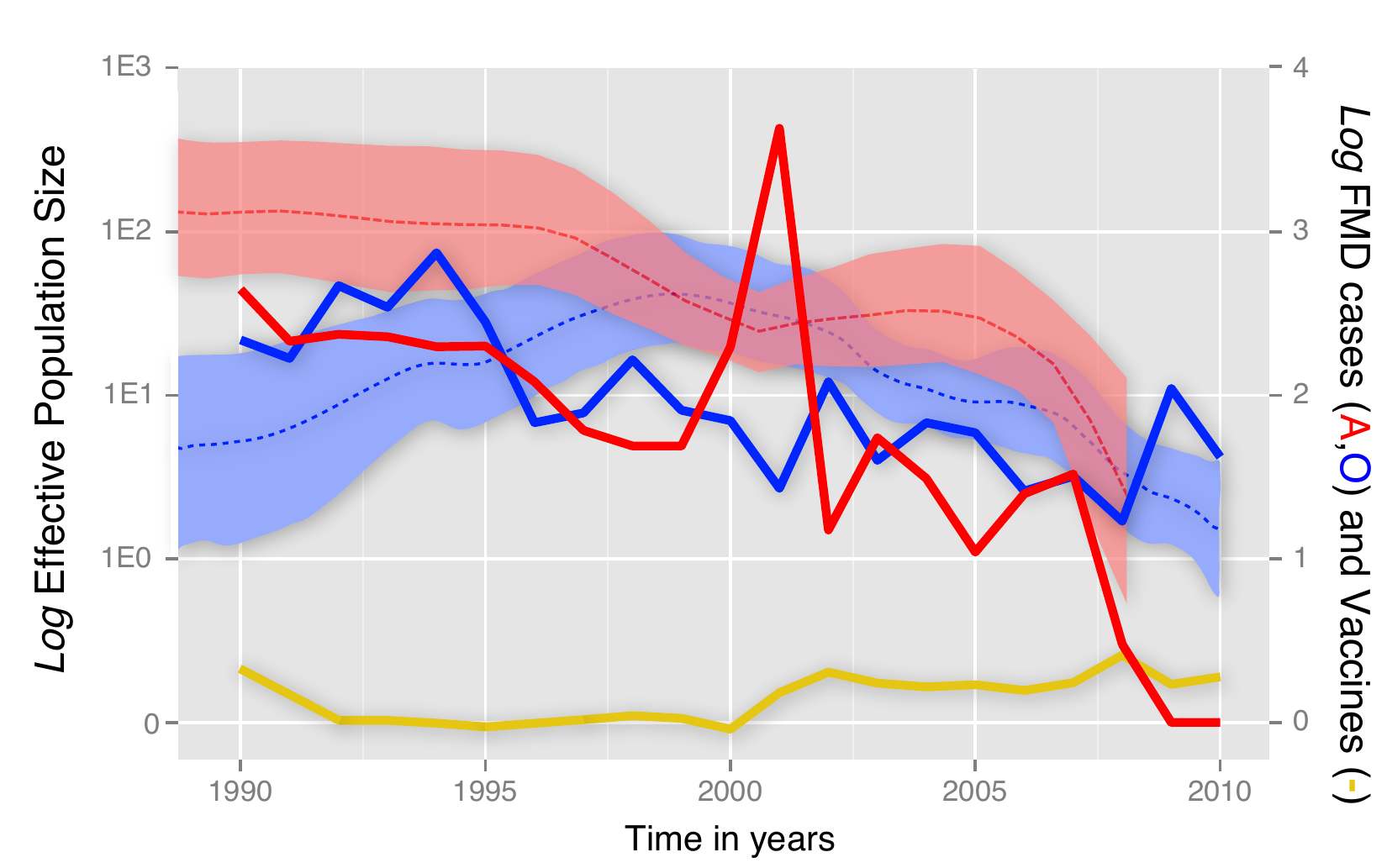}
\end{center}
\caption{}
\label{fig:skyride}
\end{figure}

The demographic reconstruction using the non-parametric skyride coalescent model shows strikingly different dynamical behavior for the two serotypes (Figure~\ref{fig:skyride}).
While serotype O exhibits a peak diversity in the late years of the 1990s, the diversity of serotype A has been slowly decreasing over the last $20$ years.
Serotype A exhibits a more stable behavior over most of the 20th century, with most variation occurring within the temporal sampling interval, mainly in the last years of the 2000s.
To gain insight into the relationship between vaccination and viral diversity, we overlay vaccination data on to the skyride plots presented by the yellow line in Figure~\ref{fig:skyride}.
These data are expressed as (log) doses per head, which we consider to be a more accurate measure of vaccination coverage, since it corrects for population size increase/decrease over time (see Figure~\ref{sfig:prod} for livestock population time series). 

Further, we overlay serotype-specific disease notification (number of cases) on the demographic reconstruction to illustrate the relationship between viral diversity and the onset of epidemics. 
For serotype A, we observe a bottleneck of viral diversity around $2001$ (red line) that coincides with a major epidemic by this serotype that affected several countries in the period $2000-2002$, especially Argentina.
The effective population size for serotype O (blue line) shows a different, more steady pattern over the years, with diversity reaching its lowest level by end of the years $2000$.
We also observe this decreasing trend in viral diversity for serotype A.
The number of vaccine doses shows a marked increase after $2001$ (note that the Y-axis in Figure~\ref{fig:skyride} in natural log units), and the number of cases also declines for both serotypes, in particular for serotype A.
Although we do not provide any formal association analysis, the general intuition is that increasing vaccination efforts as well as other preventive measures decisively decreased transmission and thus viral diversity.
Finally, to assess the robustness of our reconstructions we perform demographic reconstructions using a) only recent (sampling date $>2000$) sequences (see Text S2, Figure ~\ref{sfig:only2000sky}) and b) data sets without the over-represented locations (data not shown).
These analyses gave very similar results to those presented in Figure~\ref{fig:skyride}.

\begin{center}
 [Figure~\ref{fig:skyride} about here]
\end{center}

\section*{Discussion}

In this paper we study the spatio-temporal evolutionary dynamics of FMDV serotypes A and O in South America, using state-of-the-art Bayesian phylogenetic methods to uncover the similarities and differences between these serotypes and assess the impact of their different biology on their population dynamics.
We identify important differences in evolutionary tempo and mode between serotypes, with different countries being important for spread.
These differences in evolutionary rate magnitude and variability suggest that, although the two serotypes share the same host range and infection routes, they present rather different evolutionary dynamics in the continent, which may help explaining their different emergence patterns (see Figure 3 in Naranjo \& Cosivi, 2013~\cite{Naranjo2013}). 

For serotype A, the spread of the virus into Brazil, Colombia and Venezuela seems to have taken place in the early decades of the 20th century while the introduction into Uruguay and Paraguay seems to have taken place much later, suggesting the former as original viral reservoirs and the latter as viral importers.
The spread of the serotype O strains circulating in South America began in Colombia around $1994$ and quickly dispersed to neighbouring countries such as Ecuador and Venezuela, suggesting that the strains circulating during the 2000s may have been introduced from elsewhere.
Our analysis however indicates that the strains isolated in Colombia in $1994$ are at the root of the circulating serotype O strains in the continent (Figure~\ref{fig:epidemictracing}).
The inclusion of archival sequences from other continents would make it possible to determine whether the circulating strains are the result of sustained maintenance or the result from multiple introductions. 

We find well-supported migration paths between Venezuela, Colombia and Ecuador for both serotypes indicating an important spread pathway in the northern part of South America, with Argentina, Brazil and Uruguay forming a well-supported (BFs $> 9$) sub-network (see Figure~\ref{fig:mj&BFs}).
Interestingly, for serotype O the Markov jumps analysis shows a clear separation in two sub-networks (Figure~\ref{fig:mj&BFs}), one comprising the Southern cone and the other comprising Andean countries.
This separation suggests that viral transmission may follow different regimes in these two regions, which have different forms of livestock production~\cite{Saraiva2003,Naranjo2013}.
The link between these two sub-networks appears to be Bolivia, where viral introduction took place around $2000$, most likely from Peru (Figure~\ref{sfig:epitrac}). 
Since both Peru, Bolivia and Venezuela are considered to have achieved less than expected regarding the implementation of control measures~\cite{Naranjo2013}, a link passing through these countries is plausible.
The northern (Andean) part of South America also stands as a major diversity reservoir for FMDV, with Colombia being the main viral seeder for serotype O and Venezuela being decisively important for serotype A maintenance in its region.
It should be stressed however that while Colombia has taken the appropriate steps to controlling FMD in its territory, Venezuela still faces challenges in implementing the necessary control policies~\cite{Naranjo2013}.

Using data on trade of live cattle, pigs and sheep between South American countries as predictors of FMDV diffusion provides further evidence of different factors influencing the spread of serotypes A and O.
The trade of cattle is the most significant predictor for serotype O spread, which is compatible with the notion that these hosts are the most important for FMDV maintenance and transmission, even though sheep population sizes are on par with those of cattle~\cite{Saraiva2003}.
Distance between countries is also an important predictor of spread, a result consistent with the finding the most location-transitions occur between countries that share borders.

Previous studies have shown that occurrences caused by serotype A present longer cycles and wider epidemics, while serotype O is more prevalent with shorter disease cycles~\cite{colombiatime}.
These epidemiological features are reflected in the temporal variation observed for viral $Ne$ in both serotypes, a result obtained for other viruses as well~\cite{Bennett2010,Pybus2003}. 
The diversity bottleneck observed for serotype A in Figure~\ref{fig:skyride} is consistent with a single strain being rapidly transmitted during the epidemic that affected several countries during $2000-2002$.
Combined with the results from the epidemic tracing presented in Figures~\ref{fig:epidemictracing} and~\ref{sfig:epitrac}, which show that Argentina most likely seeded the outbreaks in Brazil and Uruguay, the demographic histories presented here provide additional evidence of trans-border diffusion as an important factor driving re-emergence in previously FMDV-free areas, as were Argentina and Uruguay at the time, for example.
The demographic reconstruction for serotype O does not show any bottlenecks, suggesting a different epidemiological scenario, in which viral introductions lead to establishment of endemicity and increased viral diversity.
Our results suggest that this serotype O lineage was introduced in Ecuador from Colombia (Figure~\ref{sfig:epitrac}E) and then underwent endemic circulation.

Overall, both spatial and temporal analyses point towards serotype O circulation in South America being characterized by endemic establishment with smaller epidemics and increased viral persistence. 
We find that Colombia was the most probable location of origin for serotype O, a result consistent with Carvalho et al. ~\cite{Carvalho2013}, who showed that a province close to the border with Colombia, where an annual animal fair takes place, was the most probable spatial origin of the strains circulating in Ecuador from $2002$ to $2010$.
Serotype A on the other hand shows lower within country diversity and seems to occur in bursts, marked by rapid trans-border spread and larger outbreaks. 
Despite these important differences, from the temporal reconstructions for both serotypes it can be deduced that over time, with the increase of vaccination coverage, viral effective population size ($Ne$) decreases dramatically, a result previously obtained for serotype O in Ecuador~\cite{Carvalho2013}.
Vaccination seems to be disrupting viral diversity, likely by precluding spread over large spatial extents, inducing a state of focalised transmission.
Since our results suggest that these two serotypes present rather different evolutionary dynamics, the overall decrease in viral diversity detected for both serotypes points towards a progressive success of the eradication program in slowly reducing transmission and viral diversity.

\section*{Methods}

\subsection*{Genetic and epidemiological data}

To study the spatio-temporal spread dynamics of FMDV within South America, we have compiled the largest database of 1D (VP1) gene sequences to date for serotypes A and O.
We have retrieved all 1D (VP1) nucleotide sequences available from the National Center for Biotechnology Information (NCBI, \url{ http://www.ncbi.nlm.nih.gov/}) for which information on country and year of isolation was available.
This resulted in 131 sequences (from eight countries) for serotype A and 167 sequences (from nine countries) for serotype O, covering time spans of 55 (1955-2008) and 16 (1994-2010) years, respectively (see Text S1 for details).
We aligned each data set using the MUSCLE~\cite{muscle} algorithm implemented in the MEGA5~\cite{MEGA} package.

Data on animal trade were obtained from the FAO database (\url{http://faostat.fao.org/}).
We retrieved data on the detailed trade matrix for cattle, pigs and sheep (number of live animals exchanged) covering the period from 1986 to 2009, for each of the nine countries.
Serotype-specific outbreak notifications were obtained from FMD Bioportal (\url{http://fmdbioportal.ucdavis.edu:8080/}).

\subsubsection*{Data availability}
All the data used in this paper, as well as code to produce many of the plots/analyses and BEAST XML files are hosted at \url{https://github.com/maxbiostat/FMDV_AMERICA}.

\subsection*{Phylogenetic Analysis}

We have checked both data sets for recombination using the SBOP and GARD~\cite{sbpgard} tools available from the Datamonkey facility (\url{http://www.datamonkey.org/}), which did not yield any indications of recombination being present in our data sets.
For all our analyses, we assume a general time reversible (GTR)~\cite{Tavare1986} model of sequence evolution, along with gamma-distributed rate heterogeneity (4 categories).
We take a Bayesian approach to testing evolutionary hypotheses while accommodating phylogenetic uncertainty. 
To this end, we use the Bayesian Evolutionary Analysis by Sampling Trees (BEAST)~\cite{beast2012} software package to infer time-structured phylogenies for the two serotypes taking advantage of the BEAGLE~\cite{BEAGLE} library to gain computational efficiency.
To compare the performance of several combinations of tree priors and molecular clocks for each data set (see Text S2), we use state-of-the-art marginal likelihood estimators, such as path sampling (PS)~\cite{LartillotPhilippe} and stepping-stone sampling (SS)~\cite{Xie}, which have only recently been introduced in the field of phylogenetics~\cite{LartillotPhilippe, Xie, Baele2012, Baele2013a, Baele2013b, Baele2013c}.

\subsection*{Quantifying temporal and spatial signal} 

To assess the temporal signal for each serotype, we use the approach described in Faria et al.~\cite{Faria2012} and Baele et. al~\cite{Baele2012} and compare the marginal likelihoods of a dated-tips model and a contemporaneous-tips model by calculating Bayes Factors (BF)~\cite{Suchard2001, suchard2005models} (see Spatial Model Selection for details).
We follow Kass and Raftery (1995)~\cite{KassRaftery1995} and consider a log BF$>$3 to be indicative of decisive support for the hypothesis of temporal structure.

We quantify spatial signal using Bayesian tip-association tests, implemented through the BaTS software package~\cite{bats}.
To detect phylogeny-location association, we assign each sequence to its country of origin and compute association index (AI) and parsimony score (PI) using BaTS on a subset of 1000 samples from the posterior distribution of topologies.
We obtain a null distribution for each statistic (AI and PI), against which the observed indices are compared and significance is assessed.
Additionally, we compute the monophyletic clade (MC) size for each state (country), as a local indicator of phylogeny-trait association for each state (country).
Please see Text S2 for further details on the BaTs analyses.

\subsection*{Spatio-temporal Dynamics}

We apply the non-parametric skyride coalescent model~\cite{skyride} in order to reconstruct the past population dynamics for both serotypes, using a Gaussian Markov Random Field (GMRF) prior to obtain smooth estimates for effective population size trajectories over time.
To gain insight into the mechanisms driving viral dynamics, we overlay the demographic reconstructions to serotype-specific outbreak and vaccination (doses per head) time series.
We perform phylogeographic analyses of FMDV in South America using the methods presented in Lemey et al.~\cite{roots}, available in BEAST, and apply an asymmetric, non-reversible discrete phylogeographic model to both data sets, with each country used as a discrete state.
For statistical efficiency, we use Bayesian stochastic search variable selection (BSSVS) to choose the minimal set of dispersal rates that sufficiently explain the observed data.
BSSVS naturally allows for assessing the significance of each migration route through a Bayes factor (BF) test.
We use SPREAD~\cite{spread} to generate KML files (visualized using Google Earth, \url{http://www.google.com/earth/index.html}) and calculate BFs for statistically significant epidemiological links between discrete locations.
We go on to compute the expected number of transitions between each pair of locations conditional on the observed data and summarize the number of transitions between countries using a robust counting approach~\cite{Minin2008}.
For these analyses we use a sample of $10,000$ trees from the posterior distribution of topologies.
Unbalanced sampling can have an important impact on the inference of the spatial migration rates~\cite{Faria2012, Lemey2014, Frost2014}.
In this study, both data sets analyzed present highly preferential sampling, with Ecuadorian sequences representing about 50\% of the serotype O data and about 45\% of serotype A sequences being from Argentina (see Text S1 for details).
We therefore conduct an extensive sensitivity analysis, exploring various sampling schemes and comparing the obtained parameter estimates.
Supplementary Text S2 shows a complete description of the parameter estimates under different sampling schemes. 

\subsubsection*{Spatial Model Selection}

In this study we exploit recent developments in Bayesian model selection~\cite{Baele2012, Baele2013a, Baele2013b, Baele2013c}, as implemented in the BEAST software program~\cite{beast2012} to compare the different epidemiological predictors.
Specifically, we perform accurate estimation of the (log) marginal likelihood using path sampling (PS)~\cite{LartillotPhilippe} and stepping-stone sampling (SS)~\cite{Xie}, two computationally demanding approaches that yield accurate estimates of model fit while accommodating phylogenetic uncertainty.
Using these (log) marginal likelihoods, it's possible to calculate Bayes Factors, which provide a measure of the relative performance of each model. 
We have estimated all the (log) marginal likelihoods in this study using 64 power posteriors, which were each run for 2 million iterations, taking up to 4 weeks of (wall time) computation for each model under evaluation. 
Using PS and SS, we first compare different demographic priors and clock models (see Supplementary Text S2) for both serotypes. 
For more details on these model selection procedures please see Supplementary Text S2.

To test the influence of different epidemiological predictors on viral diffusion through space, we use trade of live cattle, pigs and sheep to parameterize priors for the CTMC rate matrix.
We normalize the numbers of live animals exchanged between countries to a mean and coefficient of variation of $1$ and use these as prior expectations in BEAST (see~\cite{roots}, pg. 14 and the Appendix in~\cite{Carvalho2013}).  
We compare these predictors to a distance-informed prior~\cite{roots}, which represents a scenario where flow occurs as a function of the inverse of the geographic distances between locations.
Finally, we compare all predictors against an equal-rates gamma prior, which considers~\textit{a priori} a scenario where there is no preferential spread among different countries~\cite{Nelson2011}.

\section*{Acknowledgments}
The authors would like to thank Ant\^onio Mendes (PANAFTOSA) for clarifications regarding the vaccination data, Matthew Hall (Edinburgh) and Oliver Pybus (Oxford) for insightful contributions and Miguel Carvalho, Felipe Figueiredo (PROCC) and Mauricio Oliveira (UFRJ) for operational support.
We acknowledge the support of the National Evolutionary Synthesis Center (NESCent) through a working group (Software for Bayesian Evolutionary Analysis).

\emph{Funding:} The research leading to these results has received funding from the European Union Seventh Framework Programme [FP7/2007-2013] under Grant Agreement no. 278433-PREDEMICS and ERC Grant agreement no. 260864.
This work was also supported by National Institutes of Health grants R01 HG006139 and National Science Foundation grants DMS 1264153.

\emph{Conflict of Interest:} none declared

\newpage
\bibliography{FMDV_AMERICA}

\begin{thebibliography}{10}
\providecommand{\url}[1]{\texttt{#1}}
\providecommand{\urlprefix}{URL }
\expandafter\ifx\csname urlstyle\endcsname\relax
  \providecommand{\doi}[1]{doi:\discretionary{}{}{}#1}\else
  \providecommand{\doi}{doi:\discretionary{}{}{}\begingroup
  \urlstyle{rm}\Url}\fi
\providecommand{\bibAnnoteFile}[1]{%
  \IfFileExists{#1}{\begin{quotation}\noindent\textsc{Key:} #1\\
  \textsc{Annotation:}\ \input{#1}\end{quotation}}{}}
\providecommand{\bibAnnote}[2]{%
  \begin{quotation}\noindent\textsc{Key:} #1\\
  \textsc{Annotation:}\ #2\end{quotation}}
\providecommand{\eprint}[2][]{\url{#2}}

\bibitem{review}
Grubman MJ, Baxt B (2004) {{F}oot-and-mouth disease}.
\newblock Clin Microbiol Rev 17: 465--493.
\bibAnnoteFile{review}

\bibitem{Perez2001}
Perez AM, Konig G, Spath E, Thurmond MC (2008) {{V}ariation in the {V}{P}1 gene
  of foot-and-mouth disease virus serotype {A} associated with epidemiological
  characteristics of outbreaks in the 2001 epizootic in {A}rgentina}.
\newblock J Vet Diagn Invest 20: 433--439.
\bibAnnoteFile{Perez2001}

\bibitem{Malirat2012}
Malirat V, Bergmann IE, de~Mendonca~Campos R, Conde F, Quiroga JL, et~al.
  (2012) {{M}olecular epidemiology of foot-and-mouth disease virus type {A} in
  {S}outh {A}merica}.
\newblock Vet Microbiol 158: 82--94.
\bibAnnoteFile{Malirat2012}

\bibitem{andean}
Malirat V, Bergmann IE, Campos RdeM, Salgado G, Sanchez C, et~al. (2011)
  {{P}hylogenetic analysis of {F}oot-and-{M}outh {D}isease {V}irus type {O}
  circulating in the {A}ndean region of {S}outh {A}merica during 2002-2008}.
\newblock Vet Microbiol 152: 74--87.
\bibAnnoteFile{andean}

\bibitem{review_eradication}
Correa~Melo E, Saraiva V, Astudillo V (2002) {{R}eview of the status of foot
  and mouth disease in countries of {S}outh {A}merica and approaches to control
  and eradication}.
\newblock Rev - Off Int Epizoot 21: 429--436.
\bibAnnoteFile{review_eradication}

\bibitem{Naranjo2013}
Naranjo J, Cosivi O (2013) {{E}limination of foot-and-mouth disease in {S}outh
  {A}merica: lessons and challenges}.
\newblock Philos Trans R Soc Lond, B, Biol Sci 368: 20120381.
\bibAnnoteFile{Naranjo2013}

\bibitem{tully}
Tully DC, Fares MA (2008) {{T}he tale of a modern animal plague: tracing the
  evolutionary history and determining the time-scale for foot and mouth
  disease virus}.
\newblock Virology 382: 250--256.
\bibAnnoteFile{tully}

\bibitem{Saraiva2003}
Saraiva V (2003) Epidemiology of {F}oot-and-mouth disease in {S}outh {A}merica.
\newblock In: Dodet B, Vicari M, editors, {F}oot and mouth disease: control
  strategies, Paris: Elsevier SAS. pp. 43--54.
\bibAnnoteFile{Saraiva2003}

\bibitem{vaccinationSA}
Saraiva V, Darsie G (2004) {{T}he use of vaccines in {S}outh {A}merican
  foot-and-mouth disease eradication programmes}.
\newblock Dev Biol (Basel) 119: 33--40.
\bibAnnoteFile{vaccinationSA}

\bibitem{cottam2007}
Cottam EM, Wadsworth J, Shaw AE, Rowlands RJ, Goatley L, et~al. (2008)
  {{T}ransmission pathways of foot-and-mouth disease virus in the {U}nited
  {K}ingdom in 2007}.
\newblock PLoS Pathog 4: e1000050.
\bibAnnoteFile{cottam2007}

\bibitem{cottam2008}
Cottam EM, Thebaud G, Wadsworth J, Gloster J, Mansley L, et~al. (2008)
  {{I}ntegrating genetic and epidemiological data to determine transmission
  pathways of foot-and-mouth disease virus}.
\newblock Proc Biol Sci 275: 887--895.
\bibAnnoteFile{cottam2008}

\bibitem{combining}
Di~Nardo A, Knowles NJ, Paton DJ (2011) {{C}ombining livestock trade patterns
  with phylogenetics to help understand the spread of foot and mouth disease in
  sub-{S}aharan {A}frica, the {M}iddle {E}ast and {S}outheast {A}sia}.
\newblock Rev - Off Int Epizoot 30: 63--85.
\bibAnnoteFile{combining}

\bibitem{Malirat2007}
Malirat V, de~Barros JJ, Bergmann IE, Campos RdeM, Neitzert E, et~al. (2007)
  {{P}hylogenetic analysis of foot-and-mouth disease virus type {O} re-emerging
  in free areas of {S}outh {A}merica}.
\newblock Virus Res 124: 22--28.
\bibAnnoteFile{Malirat2007}

\bibitem{Malirat2011}
Malirat V, Bergmann IE, Campos RdeM, Salgado G, Sanchez C, et~al. (2011)
  {{P}hylogenetic analysis of {F}oot-and-{M}outh {D}isease {V}irus type {O}
  circulating in the {A}ndean region of {S}outh {A}merica during 2002-2008}.
\newblock Vet Microbiol 152: 74--87.
\bibAnnoteFile{Malirat2011}

\bibitem{Maradei2013}
Maradei E, Malirat V, Beascoechea CP, Benitez EO, Pedemonte A, et~al. (2013)
  {{C}haracterization of a type {O} foot-and-mouth disease virus re-emerging in
  the year 2011 in free areas of the {S}outhern {C}one of {S}outh {A}merica and
  cross-protection studies with the vaccine strain in use in the region}.
\newblock Vet Microbiol 162: 479--490.
\bibAnnoteFile{Maradei2013}

\bibitem{grenfell}
Grenfell BT, Pybus OG, Gog JR, Wood JL, Daly JM, et~al. (2004) {{U}nifying the
  epidemiological and evolutionary dynamics of pathogens}.
\newblock Science 303: 327--332.
\bibAnnoteFile{grenfell}

\bibitem{vphylodynamics}
Volz EM, Koelle K, Bedford T (2013) {{V}iral phylodynamics}.
\newblock PLoS Comput Biol 9: e1002947.
\bibAnnoteFile{vphylodynamics}

\bibitem{roots}
Lemey P, Rambaut A, Drummond AJ, Suchard MA (2009) {{B}ayesian phylogeography
  finds its roots}.
\newblock PLoS Comput Biol 5: e1000520.
\bibAnnoteFile{roots}

\bibitem{movements}
Fevre EM, Bronsvoort BM, Hamilton KA, Cleaveland S (2006) {{A}nimal movements
  and the spread of infectious diseases}.
\newblock Trends Microbiol 14: 125--131.
\bibAnnoteFile{movements}

\bibitem{Nelson2011}
Nelson MI, Lemey P, Tan Y, Vincent A, Lam TT, et~al. (2011) {{S}patial dynamics
  of human-origin {H}1 influenza {A} virus in {N}orth {A}merican swine}.
\newblock PLoS Pathog 7: e1002077.
\bibAnnoteFile{Nelson2011}

\bibitem{towards}
Faria NR, Suchard MA, Rambaut A, Lemey P (2011) {{T}oward a quantitative
  understanding of viral phylogeography}.
\newblock Curr Opin Virol 1: 423--429.
\bibAnnoteFile{towards}

\bibitem{beast2012}
Drummond AJ, Suchard MA, Xie D, Rambaut A (2012) Bayesian phylogenetics with
  beauti and the beast 1.7.
\newblock Molecular biology and evolution 29: 1969--1973.
\bibAnnoteFile{beast2012}

\bibitem{phyml}
Guindon S, Gascuel O (2003) {{A} simple, fast, and accurate algorithm to
  estimate large phylogenies by maximum likelihood}.
\newblock Syst Biol 52: 696--704.
\bibAnnoteFile{phyml}

\bibitem{Faria2012}
Faria NR, Hodges-Mameletzis I, Silva JC, Rodes B, Erasmus S, et~al. (2012)
  {{P}hylogeographical footprint of colonial history in the global dispersal of
  human immunodeficiency virus type 2 group {A}}.
\newblock J Gen Virol 93: 889--899.
\bibAnnoteFile{Faria2012}

\bibitem{Baele2012}
Baele G, Lemey P, Bedford T, Rambaut A, Suchard MA, et~al. (2012) {{I}mproving
  the accuracy of demographic and molecular clock model comparison while
  accommodating phylogenetic uncertainty}.
\newblock Mol Biol Evol 29: 2157--2167.
\bibAnnoteFile{Baele2012}

\bibitem{MEP}
Drummond A, Pybus O, Rambaut A, Forsberg R, Rodrigo A (2003) {{M}easurably
  evolving populations}.
\newblock Trends in Ecology \& Evolution 18: 481-488.
\bibAnnoteFile{MEP}

\bibitem{Carvalho2013}
Carvalho LM, Santos LB, Faria NR, de~Castro~Silveira W (2013) {{P}hylogeography
  of foot-and-mouth disease virus serotype {O} in {E}cuador}.
\newblock Infect Genet Evol 13: 76--88.
\bibAnnoteFile{Carvalho2013}

\bibitem{Muellner2011}
Muellner P, Zadoks RN, Perez AM, Spencer SE, Schukken YH, et~al. (2011) {{T}he
  integration of molecular tools into veterinary and spatial epidemiology}.
\newblock Spat Spatiotemporal Epidemiol 2: 159--171.
\bibAnnoteFile{Muellner2011}

\bibitem{bats}
Parker J, Rambaut A, Pybus OG (2008) {{C}orrelating viral phenotypes with
  phylogeny: accounting for phylogenetic uncertainty}.
\newblock Infect Genet Evol 8: 239--246.
\bibAnnoteFile{bats}

\bibitem{Minin2008}
Minin VN, Suchard MA (2008) {{C}ounting labeled transitions in continuous-time
  {M}arkov models of evolution}.
\newblock J Math Biol 56: 391--412.
\bibAnnoteFile{Minin2008}

\bibitem{fluPNAS}
Bahl J, Nelson MI, Chan KH, Chen R, Vijaykrishna D, et~al. (2011) {{T}emporally
  structured metapopulation dynamics and persistence of influenza {A} {H}3{N}2
  virus in humans}.
\newblock Proc Natl Acad Sci USA 108: 19359--19364.
\bibAnnoteFile{fluPNAS}

\bibitem{Bedford2010}
Bedford T, Cobey S, Beerli P, Pascual M (2010) {{G}lobal migration dynamics
  underlie evolution and persistence of human influenza {A} ({H}3{N}2)}.
\newblock PLoS Pathog 6: e1000918.
\bibAnnoteFile{Bedford2010}

\bibitem{polar}
Edwards CJ, Suchard MA, Lemey P, Welch JJ, Barnes I, et~al. (2011) {{A}ncient
  hybridization and an {I}rish origin for the modern polar bear matriline}.
\newblock Curr Biol 21: 1251--1258.
\bibAnnoteFile{polar}

\bibitem{KL}
Kullback S, Leibler R (1951) {{O}n information and sufficiency}.
\newblock Annals of Mathematical Statistics 22: 79--86.
\bibAnnoteFile{KL}

\bibitem{colombiatime}
Gallego ML, Perez AM, Thurmond MC (2007) {{T}emporal and spatial distributions
  of foot-and-mouth disease under three different strategies of control and
  eradication in {C}olombia (1982-2003)}.
\newblock Vet Res Commun 31: 819--834.
\bibAnnoteFile{colombiatime}

\bibitem{Bennett2010}
Bennett SN, Drummond AJ, Kapan DD, Suchard MA, Munoz-Jordan JL, et~al. (2010)
  {{E}pidemic dynamics revealed in dengue evolution}.
\newblock Mol Biol Evol 27: 811--818.
\bibAnnoteFile{Bennett2010}

\bibitem{Pybus2003}
Pybus OG, Drummond AJ, Nakano T, Robertson BH, Rambaut A (2003) {{T}he
  epidemiology and iatrogenic transmission of hepatitis {C} virus in {E}gypt: a
  {B}ayesian coalescent approach}.
\newblock Mol Biol Evol 20: 381--387.
\bibAnnoteFile{Pybus2003}

\bibitem{muscle}
Edgar RC (2004) {{M}{U}{S}{C}{L}{E}: multiple sequence alignment with high
  accuracy and high throughput}.
\newblock Nucleic Acids Res 32: 1792--1797.
\bibAnnoteFile{muscle}

\bibitem{MEGA}
Tamura K, Peterson D, Peterson N, Stecher G, Nei M, et~al. (2011)
  {{M}{E}{G}{A}5: molecular evolutionary genetics analysis using maximum
  likelihood, evolutionary distance, and maximum parsimony methods}.
\newblock Mol Biol Evol 28: 2731--2739.
\bibAnnoteFile{MEGA}

\bibitem{sbpgard}
Kosakovsky~Pond SL, Posada D, Gravenor MB, Woelk CH, Frost SD (2006)
  {{A}utomated phylogenetic detection of recombination using a genetic
  algorithm}.
\newblock Mol Biol Evol 23: 1891--1901.
\bibAnnoteFile{sbpgard}

\bibitem{Tavare1986}
Tavar\'{e} S (1986) {Some Probabilistic and Statistical Problems in the
  Analysis of DNA Sequences}, Amer Mathematical Society, volume~17.
\newblock pp. 57--86.
\bibAnnoteFile{Tavare1986}

\bibitem{BEAGLE}
Ayres DL, Darling A, Zwickl DJ, Beerli P, Holder MT, et~al. (2012)
  {{B}{E}{A}{G}{L}{E}: an application programming interface and
  high-performance computing library for statistical phylogenetics}.
\newblock Syst Biol 61: 170--173.
\bibAnnoteFile{BEAGLE}

\bibitem{LartillotPhilippe}
Lartillot N, Philippe H (2006) Computing {B}ayes factors using thermodynamic
  integration.
\newblock Syst Biol 55: 195--207.
\bibAnnoteFile{LartillotPhilippe}

\bibitem{Xie}
Xie W, Lewis PO, Fan Y, Kuo L, Chen MH (2011) Improving marginal likelihood
  estimation for {B}ayesian phylogenetic model selection.
\newblock Syst Biol 60: 150--160.
\bibAnnoteFile{Xie}

\bibitem{Baele2013a}
Baele G, Li WLS, Drummond AJ, Suchard MA, Lemey P (2013) {A}ccurate model
  selection of relaxed molecular clocks in bayesian phylogenetics.
\newblock Mol Biol Evol 30: 239--243.
\bibAnnoteFile{Baele2013a}

\bibitem{Baele2013b}
Baele G, Lemey P (2013) Bayesian evolutionary model testing in the
  phylogenomics era: matching model complexity with computational efficiency.
\newblock Bioinformatics 29: 1970--1979.
\bibAnnoteFile{Baele2013b}

\bibitem{Baele2013c}
Baele G, Lemey P, Vansteelandt S (2013) Make the most of your samples: Bayes
  factor estimators for high-dimensional models of sequence evolution.
\newblock BMC Bioinformatics 14: 85.
\bibAnnoteFile{Baele2013c}

\bibitem{Suchard2001}
Suchard MA, Weiss RE, Sinsheimer JS (2001) {{B}ayesian selection of
  continuous-time {M}arkov chain evolutionary models}.
\newblock Mol Biol Evol 18: 1001--1013.
\bibAnnoteFile{Suchard2001}

\bibitem{suchard2005models}
Suchard MA, Weiss RE, Sinsheimer JS (2005) Models for estimating {B}ayes
  factors with applications to phylogeny and tests of monophyly.
\newblock Biometrics 61: 665--673.
\bibAnnoteFile{suchard2005models}

\bibitem{KassRaftery1995}
Kass R, Raftery A (1995) {{B}ayes {F}actors}.
\newblock J Amer Statist Assoc 90: 773-795.
\bibAnnoteFile{KassRaftery1995}

\bibitem{skyride}
Minin VN, Bloomquist EW, Suchard MA (2008) {{S}mooth skyride through a rough
  skyline: {B}ayesian coalescent-based inference of population dynamics}.
\newblock Mol Biol Evol 25: 1459--1471.
\bibAnnoteFile{skyride}

\bibitem{spread}
Bielejec F, Rambaut A, Suchard MA, Lemey P (2011) {{S}{P}{R}{E}{A}{D}: spatial
  phylogenetic reconstruction of evolutionary dynamics}.
\newblock Bioinformatics 27: 2910--2912.
\bibAnnoteFile{spread}

\bibitem{Lemey2014}
Lemey P, Rambaut A, Bedford T, Faria N, Bielejec F, et~al. (2014) {{U}nifying
  viral genetics and human transportation data to predict the global
  transmission dynamics of human influenza {H}3{N}2}.
\newblock PLoS Pathog 10: e1003932.
\bibAnnoteFile{Lemey2014}

\bibitem{Frost2014}
Frost SD, Pybus OG, Gog JR, Viboud C, Bonhoeffer S, et~al. (2014) Eight
  challenges in phylodynamic inference.
\newblock Epidemics : -.
\bibAnnoteFile{Frost2014}

\end{thebibliography}


\begin{thebibliography}{1}
\providecommand{\url}[1]{\texttt{#1}}
\providecommand{\urlprefix}{URL }
\expandafter\ifx\csname urlstyle\endcsname\relax
  \providecommand{\doi}[1]{doi:\discretionary{}{}{}#1}\else
  \providecommand{\doi}{doi:\discretionary{}{}{}\begingroup
  \urlstyle{rm}\Url}\fi
\providecommand{\bibAnnoteFile}[1]{%
  \IfFileExists{#1}{\begin{quotation}\noindent\textsc{Key:} #1\\
  \textsc{Annotation:}\ \input{#1}\end{quotation}}{}}
\providecommand{\bibAnnote}[2]{%
  \begin{quotation}\noindent\textsc{Key:} #1\\
  \textsc{Annotation:}\ #2\end{quotation}}
\providecommand{\eprint}[2][]{\url{#2}}

\bibitem{Wang2001}
Wang TH, Donaldson YK, Brettle RP, Bell JE, Simmonds P (2001) {{I}dentification
  of shared populations of human immunodeficiency virus type 1 infecting
  microglia and tissue macrophages outside the central nervous system}.
\newblock J Virol 75: 11686--11699.
\bibAnnoteFile{Wang2001}

\bibitem{Salemi2005}
Salemi M, Lamers SL, Yu S, de~Oliveira T, Fitch WM, et~al. (2005)
  {{P}hylodynamic analysis of human immunodeficiency virus type 1 in distinct
  brain compartments provides a model for the neuropathogenesis of
  {A}{I}{D}{S}}.
\newblock J Virol 79: 11343--11352.
\bibAnnoteFile{Salemi2005}

\bibitem{Fitch1971}
Fitch WM (1971) Toward defining the course of evolution: minimum change for a
  specific tree topology.
\newblock Systematic Biology 20: 406--416.
\bibAnnoteFile{Fitch1971}

\bibitem{Newton}
Newton MA, Raftery AE (1994) Approximating {B}ayesian inference with the
  weigthed likelihood bootstrap.
\newblock J R Stat Soc B 56: 3--48.
\bibAnnoteFile{Newton}

\bibitem{PET}
Toft JSP, Jensen JJ, Philipsen P (2010) PET: Simulation and Reconstruction of
  PET Images.
\newblock \urlprefix\url{http://CRAN.R-project.org/package=PET}.
\newblock R package version 0.4.9.
\bibAnnoteFile{PET}

\end{thebibliography}
\section*{Figure Legends}

{\bf Figure~\ref{fig:trees}. Phylogenetic relationships of foot-and-mouth disease virus (FMDV) serotypes A and O isolates from South America.} 
Time-scaled phylogenetic maximum clade credibility (MCC) trees for FMDV VP1 sequences from eight countries in the period 1955-2010 for serotype A (Panel A) and nine countries 1994-2010 for serotype O (Panel B).
Tips were collapsed for clarity and colored according to geographic origin.
Diamond sizes are proportional to posterior probabilities.
From these it's clear that there is considerable interspersing of sequences from different locations within clades.
In particular, Colombian and Ecuadorian sequences form mixed clades in both phylogenies.
The tree for serotype A (panel A) shows considerably more interspersing than the one for serotype O (panel B), but overall both phylogenies suggest a considerable degree of spatial mixing.

{\bf Figure~\ref{fig:migration}. Spatio-temporal dynamics of FMDV in South America.} 
Using an asymmetric diffusion model, we reconstruct the spatial spread of FMDV serotypes A and O throughout the South American continent during the 20th century.
The figure presents spatial projections of the maximum clade credibility trees produced with SPREAD~\cite{spread} and visualized using Google Earth (\url{http://www.google.com/earth/index.html}).
Circle radiuses are proportional to lineage diversity.
For serotype A (left hand panels), some long-range migration events have taken place in the period $1945-1965$ and may have led to the establishment of viral circulation in the Southern Cone.
Contrasting to this, for serotype O dispersal from the northern part of the continent to the Southern Cone seems to have been bridged through Bolivia (panel B).
Regarding diversity, Colombia, Venezuela and Ecuador stand out as the main reservoirs of viral diversity for both serotypes.

{\bf Figure~\ref{fig:mj&BFs}. Migration networks for FMDV serotypes A and O in South America.} 
We estimate the number of migration events between countries using the robust counting (Markov jumps) techinique of~\cite{Minin2008}.
Additionally, we perform a separate analysis employing Bayesian Stochastic Search Variable Selection (BSSVS) to determine the most significant migration routes.
Bayes factors are depicted by arrows, with line thickness proportional to BF magnitude (we only plot BFs bigger than $3$).
Coroplethic maps show the net migration rates for each country, for both serotypes.
From this it's clear that the pattern of spatial spread is different for both serotypes, although there is agreement on the linkage between Venezuela and Colombia and Ecuador and Peru.
For serotype O, Bolivia stands out as a hub, while for serotype A Brazil is highly connected.
It should be noted that the exchange rates (depicted by the colors) are considerably higher for serotype A.

{\bf Figure~\ref{fig:epidemictracing}. Epidemic tracing using robust counting for serotypes A and O in South America.} 
We show the most probable sources of serotype A epidemics in Argentina 2001 (A) and Ecuador 2002 (B).
For serotype O the origins of all the Colombian sequences from 1994 (C) are shown along with the origins of the strain in Venezuela 2003 (D).
The strains in the Argentinian $2001$ outbreak were probably already circulating in the country (panel A), while the strains isolated in Ecuador $2002$ were most likely originated in Venezuela (panel B) rather then in Argentina, where a major outbreak had occurred the previous year.
For serotype O, the circulation in Colombia has most likely being occurring since before $1994$ (panel C).
In addition to being the most probable root of the circulating strains, Colombia seems also to have seeded the introduction of serotype O in Venezuela around $2003$ (panel D).

{\bf Figure~\ref{fig:skyride}. Temporal dynamics of FMDV serotypes A and O in South America.} 
We reconstruct the population dynamics for both serotypes using the Bayesian skyride demographic prior (see Methods).
Additionally, we superimpose data on vaccination (doses per head) and (log) FMD serotype-specific notifications on the demographic reconstruction, with 95 \% Bayesian credibility intervals shaded in gray.
Combined with the results from the epidemic tracing presented in Figures~\ref{fig:epidemictracing} and~\ref{sfig:epitrac}, which show that Argentina most likely seeded the outbreaks in Brazil and Uruguay, this provides additional evidence of trans-border diffusion as a factor driving re-emergence in previously FMDV-free areas.
The demographic reconstruction for serotype O does not show any bottlenecks, suggesting a different epidemiological scenario, in which viral introductions lead to establishment of endemicity and increased viral diversity .
Since our results suggest that these two serotypes present rather different evolutionary dynamics, the overall decrease in viral diversity detected for both serotypes points towards a progressive success of the eradication program in slowly reducing transmission and viral diversity.

\newpage
\section{Figures and Tables}
\newpage
\newpage
\newpage
\newpage
\newpage
\begin{table}[H]
\caption{
\textbf{Spatial model selection results for epidemiological predictors.}
We assess the significance of livestock trade in FMDV spread in South America using two marginal likelihood calculation methods, path sampling (PS) and stepping-stone sampling (SS), to estimate (log) marginal likelihoods for each predictor using $64$ path steps and $2$ million iterations per path step, and corresponding Bayes factor (BF) comparisons.
We also present the (log) marginal likelihood estimates for two location priors under which the location rates are estimated, rather than being fixed (as for the livestock predictors).
These results show that inverse-geographic distance is a strong predictor of viral spread for both serotypes.
Interestingly, while the best predictor for serotype O was the trade of cattle, this predictor fails to yield outperform the equal-rates prior.
These results provide evidence that the spread determinants of serotypes A and O are different in the continent.
}
\begin{center}
\begin{tabular}{lrrrrrr}
\toprule
 & \multicolumn{3}{c}{Serotype A}& \multicolumn{3}{c}{Serotype O}\\
 \midrule
Predictor & PS & SS & log BF$^2$ & PS & SS & log BF \\
Cattle&-12588.76&-12591.26&-27.70&\textbf{-8308.94}&\textbf{-8311.21}& \textbf{13.49}\\
Distance&\textbf{-12557.69}&\textbf{-12559.73}&\textbf{3.83}&-8313.89&-8315.37&9.33\\
Pigs&-12589.33&-12590.94&-27.38&-8325.39&-8326.63&-1.93\\
Sheep&-12570.67&-12572.56&-9.00&-8326.23&-8330.64&-5.94\\
\\
\hline
Equal rates &-12561.98&-12563.56&--&-8321.49&-8324.70&--\\
\bottomrule
\end{tabular}
\end{center}
\begin{flushleft}
\end{flushleft}
\label{tab:preds}
 \end{table}
\newpage
\begin{table}[H]
\caption{
\textbf{Inferred root locations for each predictor for both serotypes.} 
We present the most probable country of origin inferred using each predictor, with associated probabilities inside parentheses. 1-- All rates equal; 2-- Probability of being the root; 3-- Kullback-Leibler divergence.
It can be noticed that there is considerable disagreement between predictors as to which is the most probable root of the circulating strains for serotype A.
The results for serotype O in the other hand consistently point to Colombia as the source of the isolates analyzed.
We use the Kullback-Liebler divergence of the posterior distribution at root to a uniform distribution as a measure of spatial signal extraction~\cite{roots}.
By this criterion, the trade of pigs and sheep for serotypes A and O respectively, are the most efficient predictors at capturing spatial signal.
}
\begin{center}
\begin{tabular}{lcccc}
\toprule
& \multicolumn{2}{c}{Serotype A}&\multicolumn{2}{c}{Serotype O}\\
Predictor& Origin ($Pr$(root)$^2$)& KL$^3$&Origin ($Pr$(root))& KL\\
\midrule
Distance & Argentina ($0.75$)& $3.86$ & Colombia ($0.96$)& $3.76$\\
Sheep    & Brazil ($0.89$) & $3.52$ & Colombia ($0.99$)& $5.91$\\
Pigs      & Colombia ($0.99$)& $5.98$& Colombia ($0.91$)& $4.73$\\
Equal rates$^1$  & Argentina ($0.84$)& $3.78$ &Colombia  ($0.96$)& $3.20$\\
Cattle   & Peru ($0.93$)& $5.17$ & Colombia ($0.95$)& $3.81$\\
 \bottomrule
\end{tabular}
\end{center}
\begin{flushleft}
\end{flushleft}
\label{tab:roots}
 \end{table}
\end{document}